\pdfoutput=1
\documentclass[a4paper]{article}
\usepackage[T1]{fontenc}
\usepackage{color, graphicx}
\usepackage[english,USenglish]{babel}
\usepackage[babel]{csquotes}
\usepackage[sort&compress,numbers]{natbib}
\usepackage{mciteplus}
\usepackage{authblk}
\usepackage{hyperref}
\usepackage{multirow}
\hypersetup{
    colorlinks=true,
    linkcolor=blue,    
    citecolor=cyan,
}

\title{Superconductivity and magnetism in iron sulfides intercalated by metal hydroxides}
\author[1,3]{Xiuquan Zhou}
\author[2,3]{Christopher Eckberg}
\author[1,2]{Brandon Wilfong}
\author[4]{Sz-Chian Liou}
\author[1]{Hector K. Vivanco}
\author[2,3]{Johnpierre Paglione}
\author[1,3]{Efrain E. Rodriguez \thanks{efrain@umd.edu}}
\affil[1]{Department of Chemistry and Biochemistry, University of Maryland, College Park, MD 20742}
\affil[2]{Department of Physics, University of Maryland, College Park, MD 20742}
\affil[3]{Center for Nanophysics and Advanced Materials, University of Maryland, College Park, MD 20742}
\affil[4]{AIM Lab, Maryland NanoCenter, University of Maryland, College Park, MD 20742}

\begin{document}

\date{\vspace{-5ex}}

\maketitle

\renewcommand{\abstractname}{\vspace{-\baselineskip}}
\begin{abstract}

Inspired by naturally occurring sulfide minerals, we present a new family of iron-based superconductors.  A metastable form of FeS known as the mineral mackinawite forms two-dimensional sheets that can be readily intercalated by various cationic guest species.  Under hydrothermal conditions using alkali metal hydroxides, we prepare three different cation and metal hydroxide-intercalated FeS phases including (Li$_{1-x}$Fe$_x$OH)FeS, [(Na$_{1-x}$Fe$_x$)(OH)$_2$]FeS, and K$_x$Fe$_{2-y}$S$_2$.   Upon successful intercalation of the FeS layer, the superconducting critical temperature $T_c$ of mackinawite is enhanced from 5 K to 8 K for the (Li$_{1-x}$Fe$_x$OH)$^{\delta+}$ intercalate. Layered heterostructures of [(Na$_{1-x}$Fe$_x$)(OH)$_2$]FeS resemble the natural mineral tochilinite, which contains an iron square lattice interleaved with a hexagonal hydroxide lattice.  Whilst heterostructured  [(Na$_{1-x}$Fe$_x$)(OH)$_2$]FeS displays long-range magnetic ordering near 15 K, K$_x$Fe$_{2-y}$S$_2$ displays short range antiferromagnetism.  

\end{abstract}

\section*{Introduction}
\
The chemistry of iron-based superconductors has been dominated by the arsenide\cite{Kamihara2008, Bos:2008wq, Chen:2008vs, delaCruz:2008tf, Luetkens:2009ut}, selenide,\cite{Hsu2008, Kotegawa_2008, Margad_2008, Imai_2009} and telluride\cite{Hanaguri_2010, Zajdel_2010, Rodriguez2011} compounds since their discovery nearly a decade ago. Many high-temperature superconductors exhibit layered structures, and rich chemistry can be applied to modify their structures that may result in the increase of their critical temperatures ($T_c$).\cite{Attfield2011, Ozawa2008} We demonstrate that iron sulfides prepared by hydrothermal routes provide a new series of superconductors that could further elucidate the structure-property relationships across closely related phases.  Mainly, we isolate FeS layers to enhance their two-dimensional (2D) electronic character by inserting metal hydroxide spacers that also act as electron donating layers. 

The tetragonal form of FeS known as mackinawite is a metastable mineral recently shown to be superconducting with a $T_c$ near 4 K.\cite{Lai2015,Borg2016} Mackinawite FeS adopts the anti-PbO structure where FeS$_4$ tetrahedra edge-share to form 2D layers held by weak van der Waals interactions.  Consequently, these layered chalcogenides are excellent hosts for intercalation chemistry.\cite{Vivanco2016} In the selenide case, the $T_c$ can be increased from 8 K\cite{Hsu2008} to 42-44 K by intercalation of alkali metal in liquid ammonia\cite{Burrard-Lucas2013,Ying2013} or (Li$_{1-x}$Fe$_x$OH)$^{\delta+}$ under hydrothermal conditions.\cite{Lu2014,Pachmayr2015} Therefore, our goal was to extend this type of chemistry to the sulfides.

\begin{figure*}
	\centering
	\includegraphics[width=0.95\columnwidth]{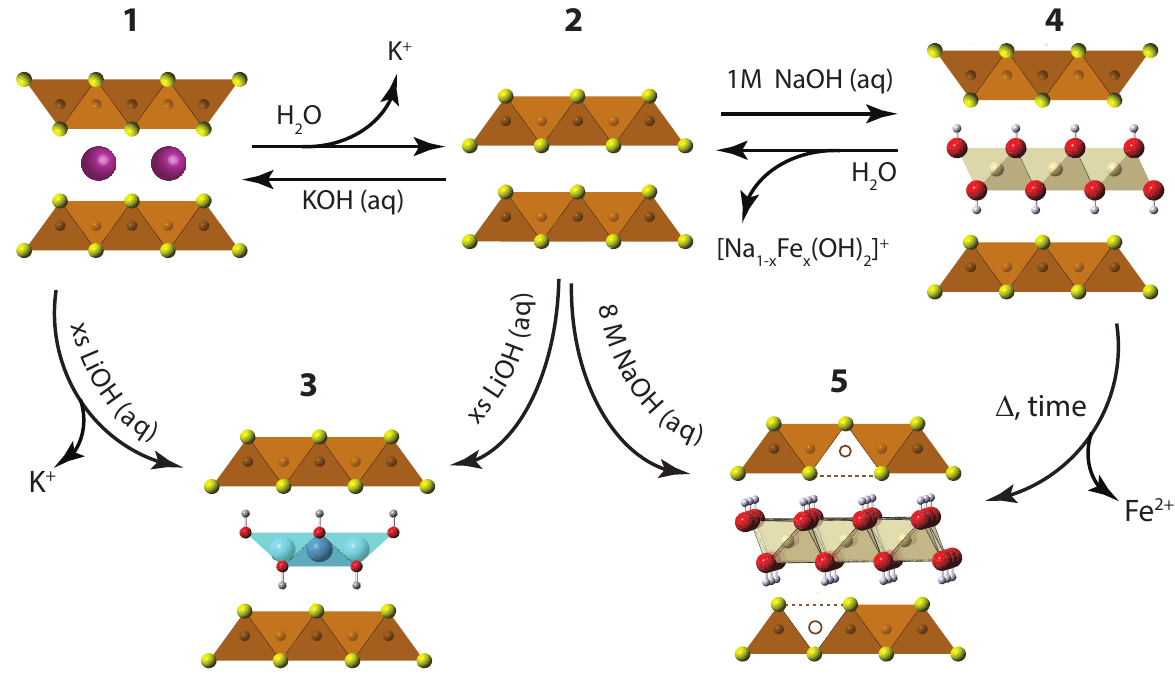}
	\caption{Synthetic scheme for the intercalation chemistry of FeS with metal hydroxides and K$^+$ cations via hydrothermal preparations.}
	\label{fig_syn}
\end{figure*}

We have found the intercalation chemistry of layered FeS to be quite versatile, and we illustrate in Fig. \ref{fig_syn} the various guest-host phases that can be prepared via hydrothermal routes.  Inspired by recent studies on the hydrothermally prepared 42 K superconductor, (Li$_{1-x}$Fe$_x$OH)FeSe,\cite{Lu2014,Pachmayr2015,Sun2015,Lynn2015,Lu:2015aa, Zhou2016} we applied similar intercalation chemistry for FeS using different alkali metal hydroxides.  Herein, we report newfound superconductivity in the Li-intercalated FeS phases, and magnetic ordering in the Na-intercalated FeS phases.  We find that the superconducting properties depend on preserving an iron square lattice and in electron doping the metallic FeS layer.

\section*{Synthesis and characterization}

For a typical preparation of (Li${_{1-x}}$Fe$_{x}$OH)FeS via the route from \textbf{2} to \textbf{3} in Fig. 1, 5 mmol Fe powder, 8 mmol of Li${_2}$S (or thiourea/Na$_{2}$S ${\cdot}$ ${9}$H$_{2}$O), 1 mmol Sn metal plate and 72 mmol LiOH${\cdot}$H$_{2}$O were mixed with 10 mL de-ionized (DI) water in a Teflon-lined stainless steel autoclave at 120-200 $^\circ$C for 3 days. Mainly Li${_2}$S was used as the sulfur source to avoid possible contamination from other alkali cations such as sodium. Afterwards, the content in the autoclave was washed and centrifuged several times until the supernatant was clear. The remaining product was collected, vacuumed dried, and stored in a nitrogen-filled glove box. 

For (Li${_{1-x}}$Fe$_{x}$OH)FeS prepared via the cation exchange route from \textbf{1} to \textbf{3} in Fig. 1, K${_x}$Fe${_{2-y}}$S${_2}$ single crystals grown from high temperature reactions were mixed with 3 mmol Fe powder, 3 mmol of sulfur source (Li${_2}$S, thiourea or Na$_{2}$S ${\cdot}$ ${9}$H$_{2}$O), 1 mmol Sn metal plate and 72 mmol LiOH${\cdot}$H$_{2}$O. The K${_x}$Fe${_{2-y}}$S${_2}$ precursors and reagents were reacted under hydrothermal conditions at 120 $^\circ$C for 1-3 days. For the growth of the K${_x}$Fe${_{2-y}}$S${_2}$ single crystals, 1.2 g of FeS powder was mixed with 0.266 g of potassium metal to match the nominal composition of KFe${_2}$S${_2}$. The FeS/K mixtures were loaded in a quartz ampoule inside a nitrogen-filled glovebox, and the ampoules flame sealed under vacuum. In order to avoid oxidation of the samples from breaking of the ampoule due to potassium-induced corrosion of the quartz walls, the sample container was sealed in a larger ampoule. For crystal growth of K${_x}$Fe${_{2-y}}$Se${_2}$, the mixture was heated to 1030 $^\circ$C over 10 h and held at 1030 $^\circ$C for 3 hours to form a homogeneous melt. Subsequently, the melt was slowly cooled at a rate of 6 $^\circ$C/hour to 650 $^\circ$C to allow crystal growth. 

For the preparation of Na-intercalated phases, we combined 10 mmol of Fe powder, 10-12 mmol of Na$_{2}$S ${\cdot}$ ${9}$H$_{2}$O, and 5-10 mmol of NaOH in an autoclave with 10 mL of DI water and heated the mixture for 7 days at 120 $^\circ$C. As described later in the Results section, these samples labeled $inc$-Na-tochilinite are compound \textbf{4} in Fig. \ref{fig_syn}. A different series of Na-intercalated samples (\textbf{5} in Fig. \ref{fig_syn}) were prepared by utilizing a larger amount of base.  The series labeled Na-tochilinite was prepared by combining 10 mmol of Fe powder, 15-20 mmol of Na$_{2}$S ${\cdot}$ ${9}$H$_{2}$O, 50-80 mmol of NaOH, and 2 mmol of Sn metal plate in an autoclave with 10 mL DI water and heated to 120 $^\circ$C for 3-7 days. 

We also utilized hydrothermal synthesis for the preparation of K-intercalated phases labeled \textbf{1} in Fig. \ref{fig_syn}. Phase pure polycrystalline material was prepared by combining 10 mmol of Fe powder, 15 mmol of thiourea, 50-100 mmol of KOH, and 2mmol of Sn metal plate with 10 mL DI water in an autoclave and heated to 160 $^\circ$C for 5-7 days.

Experimental details on the diffraction, magnetization, transport measurements, and other characterization techniques can be found in Electronic Supplemental Information (ESI) file.

\section*{Results and discussions} 


\subsection*{Li-intercalated phases}

We first describe our results utilizing LiOH to intercalate the FeS host.  Our starting point is to utilize K$_x$Fe$_{2-y}$S$_2$ (\textbf{1}) crystals grown from congruently melting the constituent elements.  Under hydrothermal and basic conditions, these crystals can either de-intercalate the potassium cations to form mackinawite FeS (\textbf{2}), or cation exchange the potassium for cationic layers of (Li$_{1-x}$Fe$_x$OH)$^{\delta+}$ as traced in the reaction from \textbf{1} to \textbf{3}.   Alternatively, we can isolate (Li$_{1-x}$Fe$_x$OH)FeS (\textbf{3}) via the method used by previous workers,\cite{Lu2014a,Pachmayr2015,Zhang2015} whereby polycrystalline material is prepared by the oxidation of iron metal in the presence of a sulfide source and excess amounts of LiOH base.  In this reaction (\textbf{2} to \textbf{3} in Fig. \ref{fig_syn}), mackinawite FeS forms in-situ with the hydroxide layers to yield (Li$_{1-x}$Fe$_x$OH)FeS.   We note that Lu \textit{et al.}\cite{Lu2014a} and Pachmayr \textit{et al.}\cite{Pachmayr2015} had previously observed superconductivity in some of their mixed solid-solutions, (Li$_{1-x}$Fe$_x$OH)FeS$_{1-z}$Se$_z$, but both studies had concluded that their pure sulfide samples ($z=0$) were nonsuperconducting. 

We found that superconductivity can be established in the intercalated sulfides for both our cation exchange and polycrystalline routes if two conditions are met: 1) the reaction temperature must be less than 160 $^{\circ}$C, i.e. mild hydrothermal conditions, and 2) the environment must remain reducing.  The latter condition was maintained by the inclusion of tin metal plate as a way to dynamically change the hydrothermal conditions from oxidizing to more reducing.\cite{Zhou2016}  No tin was found in the products as determined from energy dispersive X-ray spectroscopy (EDS).


\begin{figure}[h!]
	\centering
	\includegraphics[width=0.45\columnwidth]{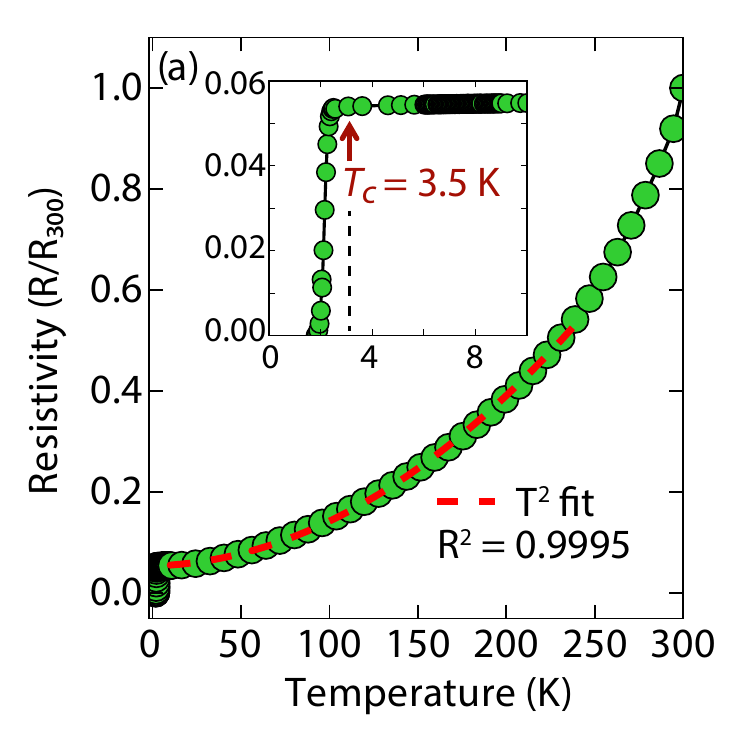}
	\includegraphics[width=0.45\columnwidth]{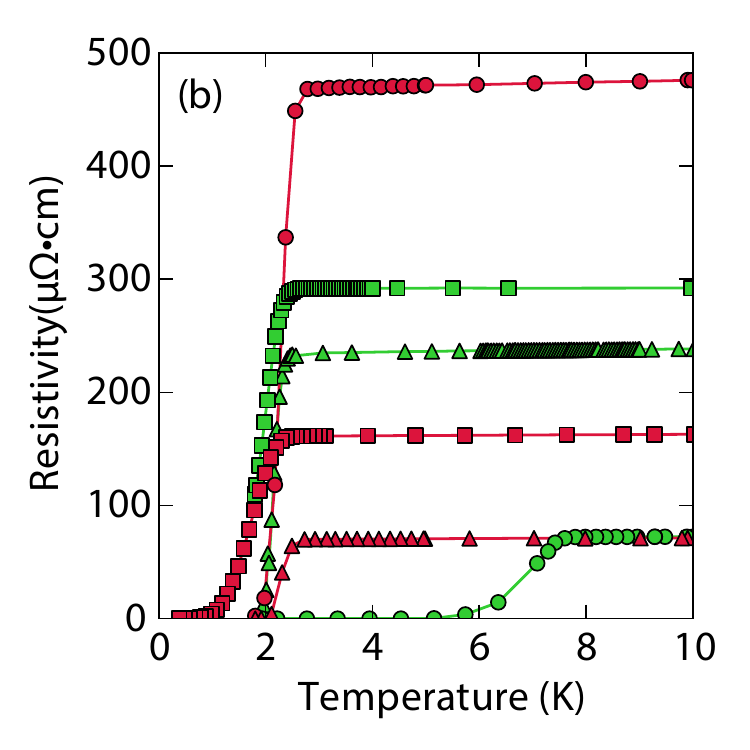}
	\caption{(a) Temperature dependent electrical resistivity of superconducting (Li$_{1-x}$Fe$_x$OH)FeS samples prepared via the cation exchange route with thiourea (b) Low temperature resistivity curves for a variety of samples prepared by either thiourea (in green) or Na$_2$S ${\cdot}$ $9$H$_2$O (in red). For (a), the $T_c$ is lower and most of the normal state resistivity (up to 250 K) can be fit with $T^2$-squared type behaviour.}
	\label{fig_PPMS_LiOH}
\end{figure}

Magnetization and electrical resistivity measurements revealed that the $T_c$ of the (Li$_{1-x}$Fe$_x$OH)FeS phases can vary from 3 K to 8 K (Fig. \ref{fig_PPMS_LiOH}), with some samples showing superconducting volume fractions up to 40\%, indicative of bulk superconductivity (Fig. S1a).  We must note, however, that due to the proximity of $T_c$ to the base temperature of our magnetometer (1.8 K) we could not reach full saturation of the diamagnetic signal.  Therefore, it is possible that the volume fraction is even higher than 40\%.  

Heat capacity measurements were also carried out in a sample with a $T_c$ near 3 K, but a large signal peaked near 4.5 K whose intensity is independent of applied magnetic field seems to mask any superconducting signal (Fig. S7a). In the similar selenides (Li$_{1-x}$Fe$_x$OH)FeSe, which have a $T_c$ near 42 K, magnetic ordering in the superconducting state takes place near 10 K due to the iron substituted for lithium in the hydroxide layer.\cite{Pachmayr2015a, Lynn2015,Lu:2015aa}  Seemingly, a magnetic signal proximate to the $T_c$ of the (Li$_{1-x}$Fe$_x$OH)FeS makes it difficult to evaluate the superconducting properties from heat capacity measurements.

Remarkably, for (Li$_{1-x}$Fe$_x$OH)FeS samples prepared via the cation exchange route, we observed $T_c$'s both above and below that of bulk FeS (Fig. \ref{fig_PPMS_LiOH}).  This result indicates that charge doping into the FeS layer is controlling the critical temperatures in (Li$_{1-x}$Fe$_x$OH)FeS.  From our various samples, intercalation by (Li$_{1-x}$Fe$_x$OH)$^{\delta+}$ could increase the $T_c$ of FeS up to 8 K.  Figure \ref{fig_PPMS_LiOH}b shows the low temperature data near $T_c$ for various samples and the sample with the lowest residual resistivity ratio also led to the highest $T_c$ in the series. From $M$ vs. $H$ hysteresis loops, the upper critical field ($H_{c2}$) of the sample at 2 K is 180 Oe whilst $H_{c1}$ was found to be approximately 40 Oe (Fig. S1b). Magnetotransport measurements find a slightly higher $H_{c2}$ near 220 Oe for $H{\parallel}c$ at 1.8 K (Fig. S2). Therefore, the intercalated compound was found to have an even smaller $H_{c2}$ than pure FeS where it is approximately 1600 Oe along the $c$-direction and 16000 along the $ab$-plane.\cite{Borg2016}  

It is also interesting to note the normal state properties of the intercalated samples.  Unlike pristine FeS,\cite{Borg2016} (Li$_{1-x}$Fe$_x$OH)FeS samples with the lower $T_c$ ($\approx$ 3.5 K) displayed nonlinear temperature dependence in the electrical resistivity above $T_c$ up to approximately 250 K, as shown by the $T^2$-fit in Fig. \ref{fig_PPMS_LiOH}a. Typically, $T^2$ dependence is associated with Fermi liquid behavior, and linear temperature dependence takes over at higher temperatures (approximately above the Debye temperature) due to electron-phonon scattering.\cite{Kittel} The samples with the lower $T_c$ exhibit this quadratic behaviour more pronouncedly (Fig. \ref{fig_PPMS_LiOH}a and Fig. S3 ). Similar Fermi liquid behaviour has been observed for the normal state in select cuprate superconductors that were overdoped in either electron and hole carriers.\cite{Levin1991, Manako1992, Cooper2009}  Another superconductor that exhibits such quadratic dependence of its resistivity near room temperature is Ag$_5$Pb$_2$O$_6$, which is a three-dimensional electron-gas system.\cite{Maeno2013}  Yonezawa and Maeno ascribe the $T^2$ behaviour to enhanced electron-electron scattering that arises in superconductors with low electron carrier densities with respect to elements such as alkali and noble metals.\cite{Maeno2013}  Therefore, it is possible that both the lower $T_c$ and $T^2$-behaviour for the sample presented in Fig. \ref{fig_PPMS_LiOH}a and Fig. S3 are related to having non-optimal charge doping in the FeS layers, and indeed lower carrier concentrations.


\begin{figure}[h!]
	\centering
	\includegraphics[width=0.85\columnwidth]{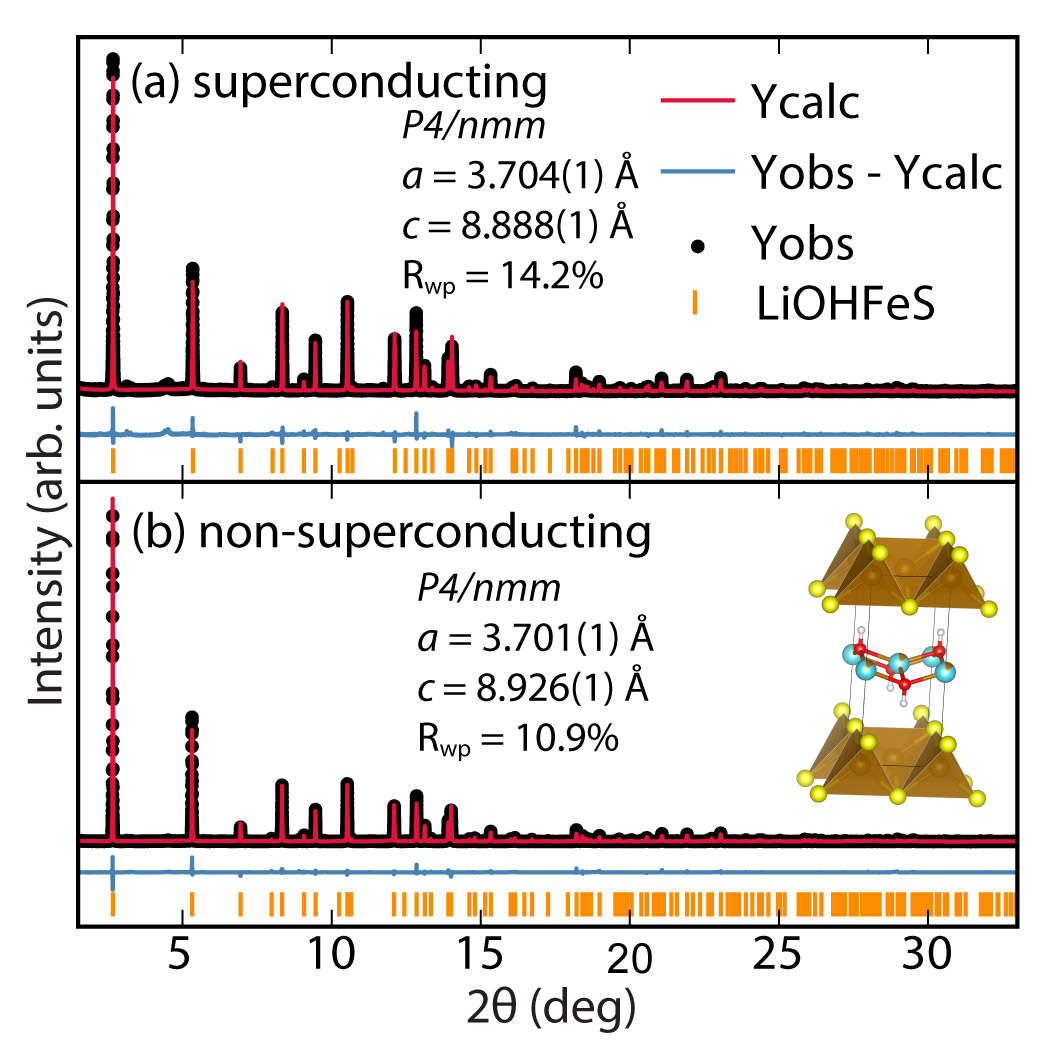}
	\caption{Synchrotron XRD patterns of (a) superconducting and (b) non-superconducting (Li$_{1-x}$Fe$_x$OH)FeS prepared under hydrothermal conditions at 160 $^{\circ}$C and 200 $^{\circ}$C, respectively. }
	\label{fig_XRD_LiOH}
\end{figure}

To determine the structure of our superconducting (Li$_{1-x}$Fe$_x$OH)FeS samples, we performed high-resolution synchrotron X-ray powder diffraction (sXRD) as shown in Fig. \ref{fig_XRD_LiOH}. From quantitative analysis of the data, we have provided detailed crystallographic information for two samples in Table \ref{Table_LiOH}. Upon intercalation, the Fe--Fe bond distances increased from 2.604 \AA \,in bulk FeS\cite{Zhou2016} to 2.619 \AA \, in (Li$_{1-x}$Fe$_x$OH)FeS, but the FeS$_4$ tetrahedron remains virtually unchanged both in bond distances and bond angles. There is also an increase in the distance between the iron square sublattices.  In mackinawite, that interlayer distance is $\approx 5.03$ \AA,\cite{Borg2016} whereas in the (Li$_{1-x}$Fe$_x$OH)-intercalated phase it is $8.89$ \AA $-8.93$ \AA, further enhancing the two-dimensionality of its electronic structure.  Due to the subtle changes in the (Li$_{1-x}$Fe$_x$OH)$^{\delta+}$ layer, Rietveld refinements for the superconducting and non-superconducting samples did not show significant differences in their stoichiometries (both close to (Li${_{0.85}}$Fe$_{0.15}$OH)FeS). 

For a more accurate analysis of chemical composition of the (Li$_{1-x}$Fe$_x$OH)FeS phases, we performed inductively coupled plasma atomic emission spectroscopy (ICP-AES). For  superconducting and non-superconducting samples, ICP-AES afforded Fe/Li ratios of 1.132 and 1.093, respectively. Since Rietveld refinements for their high-resolution synchrotron data suggested no Fe vacancy in the FeS layers (Table \ref {Table_LiOH}), the excess amounts of Fe likely resided in the LiOH layers. Therefore, the superconducting samples contains more Fe in the hydroxide layer and consequently more electron doping (0.13 e$^-$ vs 0.09 e$^-$) into the FeS layer. Similarly, Zhou \textit{et al.}\cite{Zhou2016} have reported that for the selenide analogues, higher $T_c$'s were achieved with lower reaction temperatures so that more iron cations could incorporate into the lithium hydroxide layer. Studies on the same system by Clarke \textit{et al.} demonstrated that the iron in the hydroxide layer is Fe$^{2+}$and that iron vacancies in the FeSe layer degraded the superconducting properties.\cite{Sun2015} Through the cation exchange method demonstrated here, vacancy formation in the FeS layer is less of a factor and achieving sufficient electron doping from the hydroxide layer is the bigger challenge.  We detail in the ESI (Table S1) the synthesis conditions for various superconducting and non-superconducting samples.

\begin{table}[b!]
	\caption{Lattice and structural parameters obtained from Rietveld refinement with synchrotron PXRD data collected at room temperature for a superconducting sample of (Li${_{1-x}}$Fe$_{x}$OH)FeS shown in Fig. 3a and a non-superconducting sample shown in Fig. 3b. Both samples are fitted to a $P4/nmm$ space group with 2 formula units in each unit cell (Z = 2). Relevant bond distances and bond angles are also presented for each compound. The tetrahedral angles $\alpha$$_{1}$ and $\alpha$$_{2}$ represent the S-Fe-S angles in and out of the basal plane, respectively.}
	\resizebox{\columnwidth}{!}{
		\begin{tabular}{l l l l l l l}
			\hline
			\hline
			\multicolumn{7}{c} {$a = 3.7041$(1) \AA, $c = 8.8877$(1) \AA, R$_{wp}$ = 14.27 \%, $T_{c}$ = 3 K} \\
			\hline
			\textbf{Atom}		&      \textbf{Wyckoff}  		& 	\textbf{\textit{x}}	&  \textbf{\textit{y}}		&	\textbf{\textit{z}}	&    \textbf{Occ.}		& \textbf{$U_{iso}$ (\AA$^2$)}	\\
			\textbf{}		&      \textbf{site}  		& 	\textbf{\textit{}}	&  \textbf{\textit{}}		&	\textbf{\textit{}}		&	\textbf{\textit{}}		& \textbf{}	\\
			\hline
			Li		&		2b		&	0  		&  	0 	 	&	0.5  			& 	0.848(1)  		&	0.0398(11)	\\
			Fe1	&		2b		&	0		&	0		&	0.5				&	0.152(1)		&	0.0398(11)	\\
			Fe2	&		2a		&	0.5  	&  	0.5 	&	0 				& 1				&	0.0091(2)		\\
			O		&		2c		&	0.5  	&  	0  	 	&	0.4184(3) 	& 1				&	0.0174(7)		\\
			S		&		2c		&	0   		&  	0.5  	&	0.1444(2)		& 1				&	0.0104(3)		\\
			\hline
			$\alpha$$_{1}$ ($^\circ$)	&      $\alpha$$_{2}$ ($^\circ$)			&	Fe-Fe (\AA)  		&  	Fe-S (\AA) &	F.U.    \\
			110.55(5) 	&      108.93(3)	& 2.6192(1)	& 2.2534(7) & (Li${_{0.85}}$Fe$_{0.15}$OH)FeS						\\
			{} & {} & {} & {} & {} & {} & {} \\
			\hline
			\hline
			\multicolumn{7}{c} {$a = 3.7011$(1) \AA, $c = 8.9257$(1) \AA, R$_{wp}$ = 10.91 \%, non-superconducting} \\
			\hline
			\textbf{Atom}		&      \textbf{Wyckoff}  		& 	\textbf{\textit{x}}	&  \textbf{\textit{y}}		&	\textbf{\textit{z}}	&    \textbf{Occ.}		& \textbf{$U_{iso}$ (\AA$^2$)}	\\
			\textbf{}		&      \textbf{site}  		& 	\textbf{\textit{}}	&  \textbf{\textit{}}		&	\textbf{\textit{}}		&	\textbf{\textit{}}		& \textbf{}	\\
			\hline
			Li		&		2b		&	0  		&  	0 	 	&	0.5				& 0.846(1)	&	0.0380(7)		\\
			Fe1	&		2b		&	0		&	0		&	0.5				& 0.154(1)	&	0.0380(7)		\\
			Fe2	&		2a		&	0.5		&  	0.5		&	0 				& 1			&	0.0092(1)		\\
			O		&		2c		&	0.5		&  	0  	 	&	0.4182(2)		& 1			&	0.0141(5)		\\
			S		&		2c		&	0   		&  	0.5 	&	0.1439(1)		& 1			&	0.0102(2)		\\
			\hline
			$\alpha$$_{1}$ ($^\circ$)	&      $\alpha$$_{2}$ ($^\circ$)			&	Fe-Fe (\AA)  		&  	Fe-S (\AA) &	F.U.     \\
			110.47(3 	&      108.98(2)	& 2.6171(1)	& 2.2527(4) & (Li$_{0.85}$Fe$_{0.15}$OH)FeS
			\\
			\hline
		\end{tabular}}
		\label{Table_LiOH}
	\end{table}

	\subsection*{Na-intercalated phases}
	
	Our next objective was to explore larger alkali metal hydroxides as intercalates.  Unlike LiOH, which favors a square lattice commensurate with that of mackinawite FeS, a similar structure for NaOH was not reproduced.  Instead, we found a new phase with very few reflections in the XRD powder pattern and its first peak corresponded to a $d$-spacing of 5.38 \AA. This phase is reminiscent of a natural mineral known as tochilinite, which consists of brucite-type Mg(OH)$_2$ layers between mackinawite-like FeS sheets.  Natural tochilinite is quasi-commensurate and its (001) reflection has a $d$-spacing of 10.72 \AA \,, which is close to twice our first reflection.  Therefore, if the first peak of our new phase is the (002) reflection, it would indicate that the FeS layers are stacked in a body-centered fashion.  Since we only observe (00$l$) reflections in our new phase, the square and hexagonal sheets are completely incommensurate to each other in the $ab$-plane.  Henceforth, we refer to this phase as $inc$-Na-tochilinite (\textbf{4} in Fig \ref{fig_syn}).  
	
	We found the new $inc$-Na-tochilinite to always coexist with some residual mackinawite FeS (Fig. S4). The ratio between $inc$-Na-tochilinite and mackinawite FeS was increased by using less Na$_2$S ${\cdot}$ $9$H$_2$O and decreased with prolonged ultrasonication, indicating conversion of $inc$-Na-tochilinite to FeS by de-intercalation and dissolution of the metal hydroxide layer.  The equilibrium between the two phases is indicated in the steps between $\textbf{2}$ and $\textbf{4}$ in Fig. \ref{fig_syn}. 
	
	
	\begin{figure}[h!]
		\centering
		\includegraphics[width=.45\linewidth]{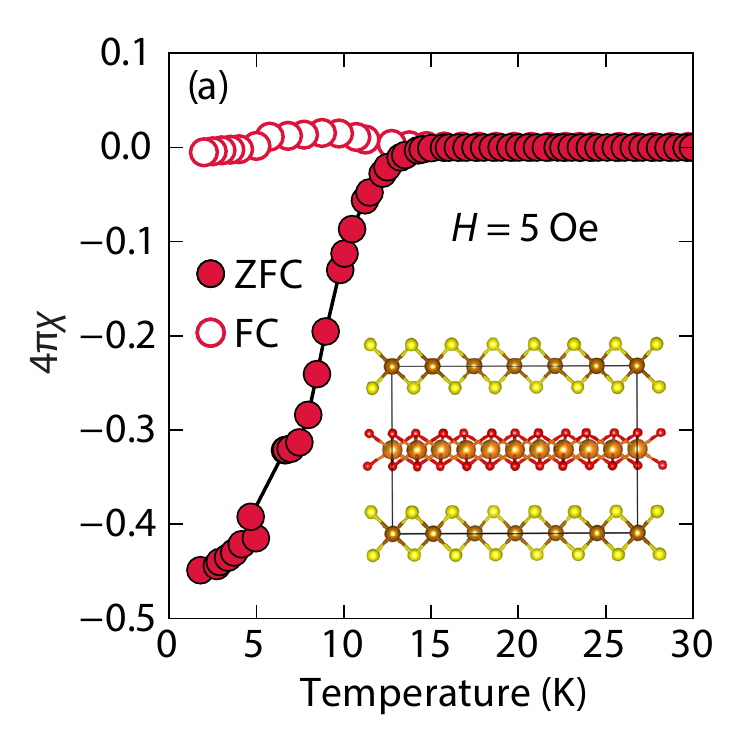}
		\includegraphics[width=.45\linewidth]{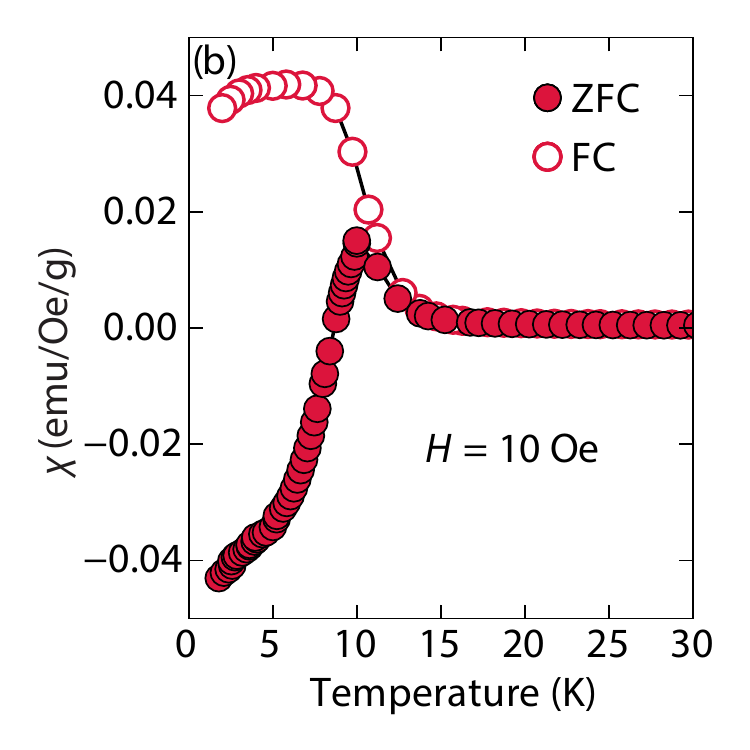}
		\caption{Magnetic susceptibility of $inc$-Na-tochilinite, [(Na$_{1-x}$Fe$_x$)(OH)$_2$]FeS, as a function of temperature with an applied external fields of (a) 5 Oe and (b) 10 Oe.}
		\label{fig_SC_NaOH}
	\end{figure}
	
	At low fields, we observed two transitions at 5 K and 15 K (Fig. \ref{fig_SC_NaOH}a).  The 5 K transition was more pronounced for a sample that contained less $inc$-Na-tochilinite and more mackinawite FeS impurity (Figs. S4 and S5 in ESI). Therefore, the 5 K anomaly likely corresponds to the superconducting transition of FeS ($T_c$ $\approx$ 4.5 K).  Although the transition at $\approx$15 K in the zero-field cooled (ZFC) curve (Fig. \ref{fig_SC_NaOH}a) appears to indicate Meissner screening due to superconductivity, the negative signal may actually correspond to long-range ordering such as ferro- or ferrimagnetism.  If the internal moment of a ferromagnetic material is of sufficient strength and aligned opposite to a weak external field, then the ZFC curve will display negative susceptibility below the Curie temperature. To resolve this ambiguity, we increased the external field of the magnetization measurements from 5 Oe to 10 Oe (Fig. \ref{fig_SC_NaOH}). The field cooled (FC) curves better indicate a clear ferromagnetic transition in $inc$-Na-tochilinite near 15 K. Therefore, $inc$-Na-tochilinite does not appear to be a superconductor based on the current magnetization data. 
	
	\begin{figure}[h!]
		\centering
		\includegraphics[width=.45\linewidth]{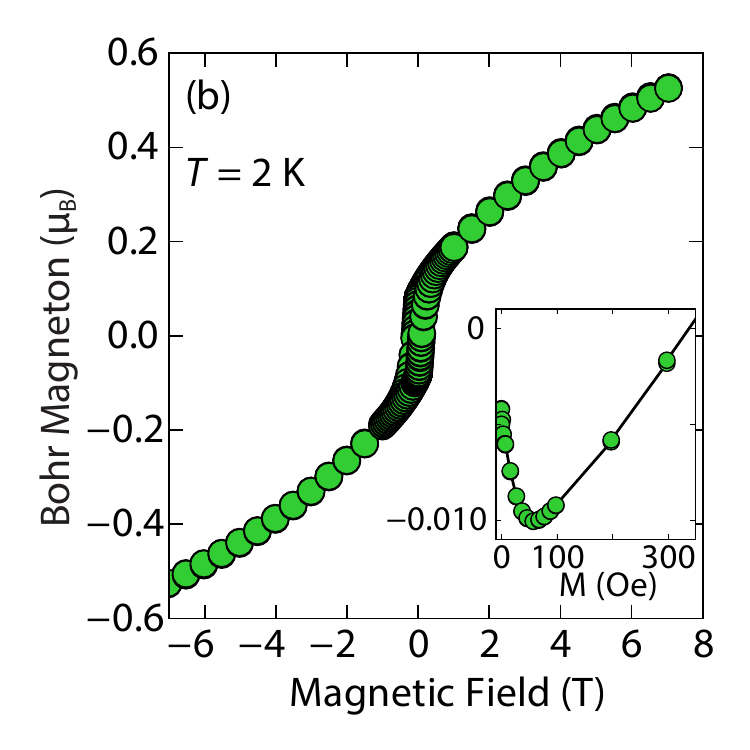}
		\includegraphics[width=.45\linewidth]{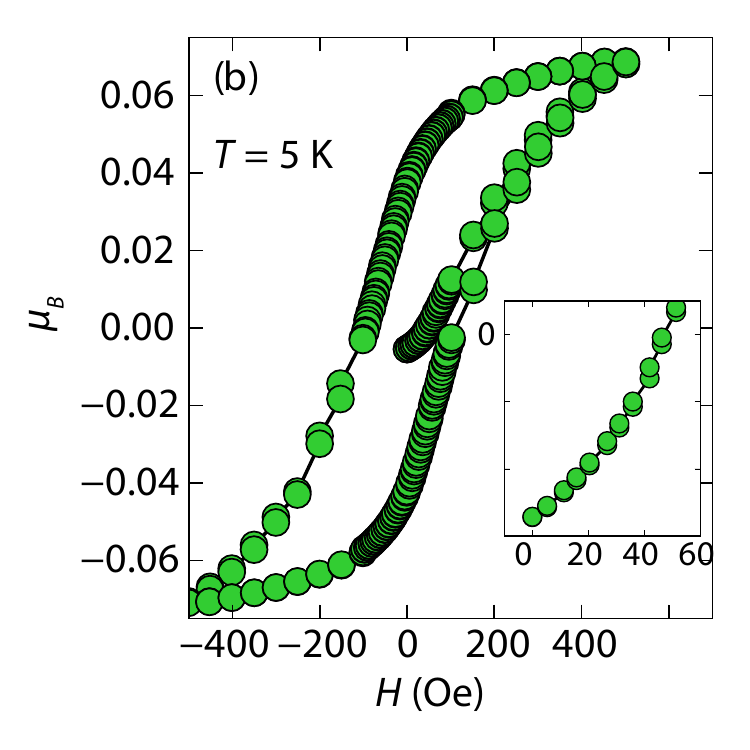}
		\caption{Magnetization versus field measurements of $inc$-Na-tochilinite, [(Na$_{1-x}$Fe$_x$)(OH)$_2$]FeS, at (a) 2 K and (b) 5 K, respectively. Inset in (a) shows the small diamagnetic region from the small amount of superconducting FeS present as an impurity phase. Inset in (b) indicates that the diamagnetic signal is lost above 5 K, which is above the $T_c$ of FeS.}
		\label{fig_NaOH_MH}
	\end{figure}
	
	We also performed temperature-dependent resistivity measurement down to 2 K on a pressed pellet of $inc$-Na-tochilinite.  We did not observe a superconducting transition, but instead semiconducting behaviour (Fig. S6). We note, however, that similar temperature-dependent behaviour was observed for pressed pellets of FeS powders,\cite{Denholme2014} even though our recent studies of of single crystal  FeS samples demonstrated that it is indeed metallic in the normal state.\cite{Borg2016} We attribute this disparity between polycrystalline and single crystal transport measurements of FeS to effects from grain boundaries and surface oxidations, typical for pressed pellets of micaceous materials. Therefore, although the current resistivity data of polycrystalline $inc$-Na-tochilinite displays semiconducting behaviour, its true state could be metallic, similar to the Li-intercalated FeS phases in the current study.     
	
	Magnetization ($M$) versus applied field ($H$) measurements further clarify the true ground state of $inc$-Na-tochilinite (Fig. \ref{fig_NaOH_MH}). The $M$ vs. $H$ curves suggest ferromagnetic behavior as the isotherm of the field sweep at 5 K (Fig. \ref{fig_NaOH_MH}b) displayed the typical hysteresis loop of ferro- and ferrimagnets. The diamagnetic signal observed for the isotherm at 2 K (Fig. \ref{fig_NaOH_MH}a inset) was therefore likely due to the superconducting FeS phase present as an impurity, which has a $T_c$ near 4.5 K.\cite{Borg2016} At 5 K, this diamagnetic signal is lost (Fig. \ref{fig_NaOH_MH}b inset). Although the new $inc$-Na-tochilinite phase is likely to be either ferro- or ferrimagnetic below 15 K, it does exhibit other interesting anomalies. The low temperature transition likely due to long-range magnetic ordering did not appear as a well defined transition in the heat capacity measurements (Fig. S7b). Instead, a broad anomaly occurred below 15 K, which was suppressed with a field of 3 T.

	\begin{figure}[t!]
		\centering
		\includegraphics[width=0.9\columnwidth]{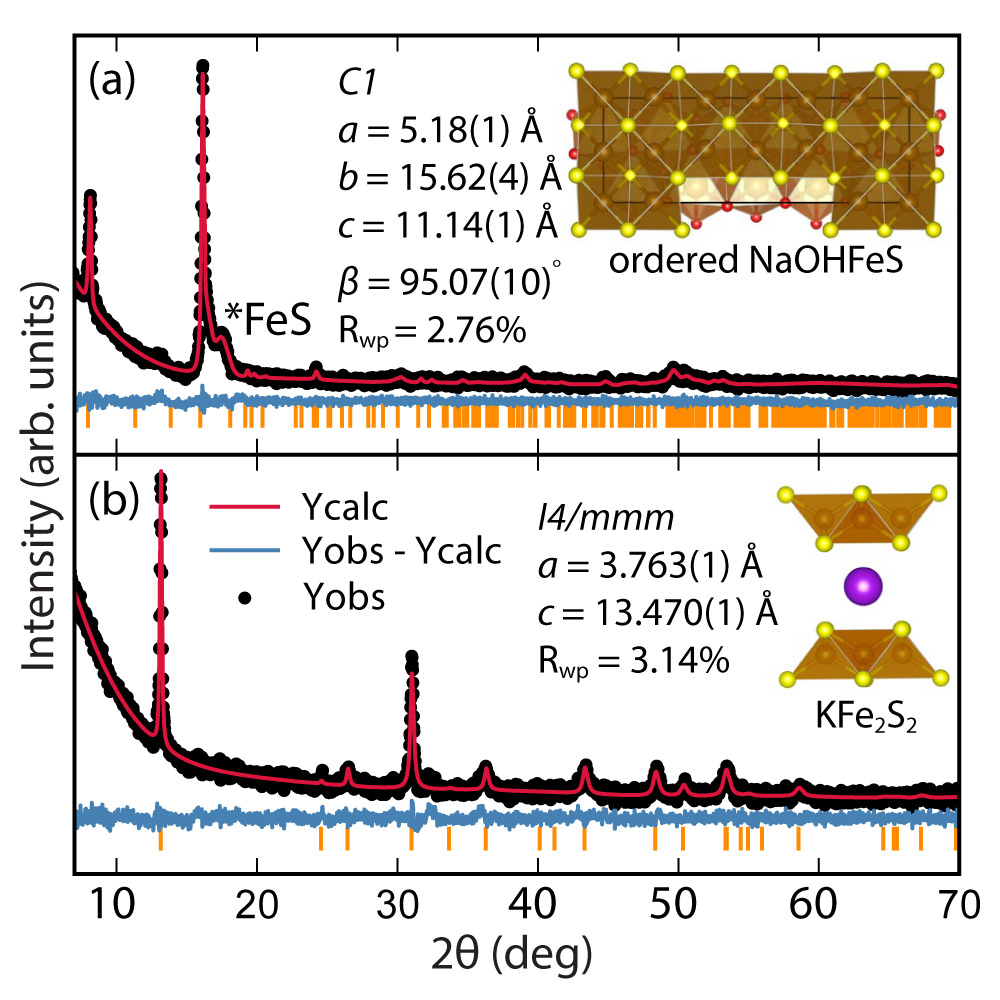}
		\caption{(a) Pawley fit to the XRD pattern of hydrothermally prepared Na-tochilinite and (b) Rietveld fit to the XRD data of K$_x$Fe$_{2-y}$S$_2$. }
		\label{fig_XRD_NaKOH}
	\end{figure}
	
	By changing the synthesis conditions of the hydrothermal reactions, the NaOH-intercalated FeS system can be stabilized into a quasi-commensurate tochilinite phase (Fig. \ref{fig_XRD_NaKOH}a), which we refer to as Na-tochilinite.  This quasi-commensurate phase was prepared by utilizing more concentrated solutions of NaOH (5 to 8 M) in the hydrothermal reactions. Significantly less tetragonal FeS was recovered (Fig. \ref{fig_XRD_NaKOH}a) with Na-tochilinite, and this phase did not easily revert to FeS by ultrasonication, indicating its stability with respect to $inc$-Na-tochilinite.  Using the crystal structure of the naturally occurring mineral known as ferrotochilinite (2(Fe${_{1-x}}$S)${\cdot}$1.8[(Mg, Fe)(OH)$_2$]),\cite{Organova1972} with lattice parameters, $a$ = 5.37 \AA, $b$ = 15.65 \AA, $c$ = 10.72 \AA, we extracted by Pawley fits the lattice parameters of our Na-tochilinite (Fig. \ref{fig_XRD_NaKOH}a)  The lattice parameters after convergence were found to be $a$ = 5.18(1) \AA \, $b$ = 15.62(4) \AA, $c$ = 11.14(4) \AA \, and $\beta$ = 95.07(10)$^{\circ}$ at room temperature.
	
\begin{figure}[t!]
		\centering
		\includegraphics[width=0.9\columnwidth]{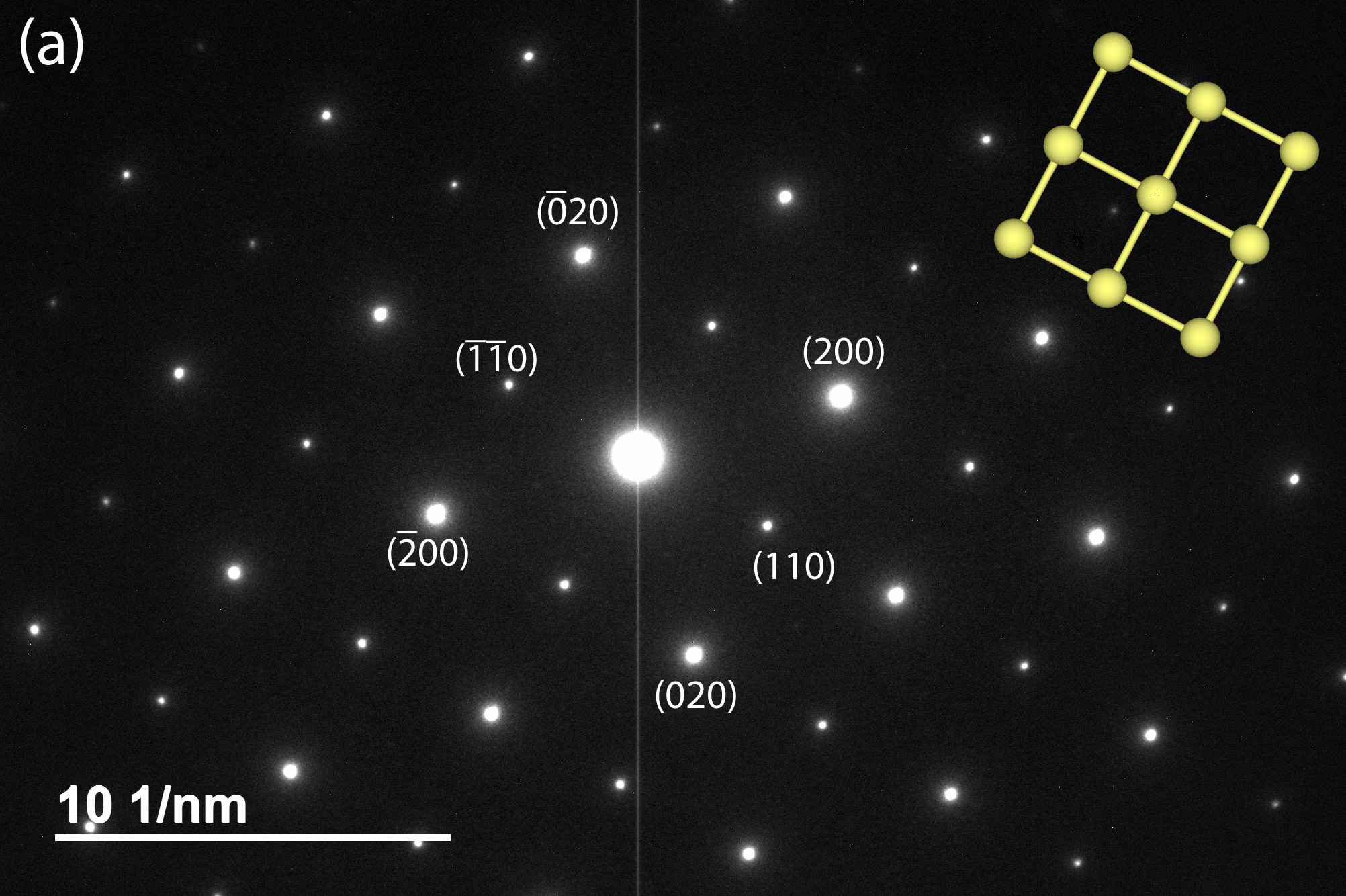}
		\includegraphics[width=0.9\columnwidth]{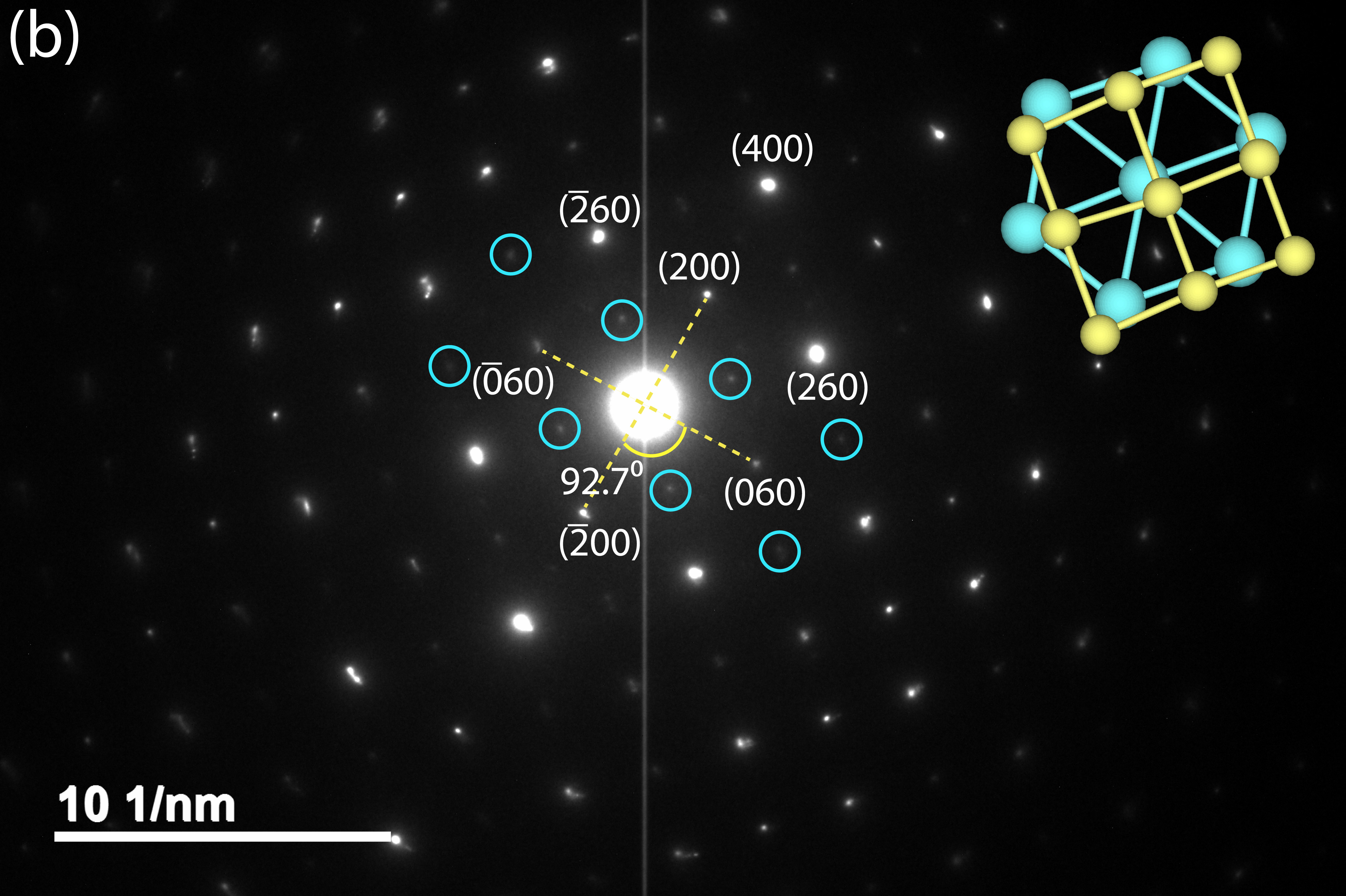}
		\caption{Electron diffraction patterns along the zone axis [001] of (a) FeS and (b) Na-tochilinite, respectively. Some weak diffraction spots of Na-Tochilinite are highlighted by blue circles. Projections of tetragonal and hexagonal lattices are shown in yellow and blue, respectively. }
		\label{fig_ED}
	\end{figure}

	Given the difficulty in elucidating the structure of these heterolayered materials by powder XRD, we have also performed electron diffraction (ED).  We present two ED patterns with the ($hk0$) reflections that were difficult to resolve from powder XRD--one for mackinawite FeS and the other for Na-tochilinite.  Along the [001] zone axis, the ED pattern of FeS (Fig. \ref{fig_ED}a) clearly demonstrates a square lattice corresponding to its simple primitive tetragonal structure.  For Na-tochilinite, additional satellite reflections emerge in addition to the square lattice of FeS (Fig. \ref{fig_ED}b). Upon closer inspection the seemingly 4-fold symmetry of the brighter reflections in Na-tochilinite is actually a 2-fold axis. The angle between the cross-sections connecting the (200) to ($\bar{2}$00) and (060) to (0$\bar{6}$0) reflections is about 93$^\circ$, which is close to the monoclinic angle found from XRD ($\beta$ = 95.07(10)$^\circ$). Therefore, the monoclinic distortion of the FeS square lattice in Na-tochilinite is clearly reproduced in the ED along with the satellite reflections indicating the intercalation of the FeS layers. The lattice constants $a$ and $b$ extracted from ED are 5.2(2) \AA \, and 15.9(2) \AA, respectively--in good agreement with the XRD analysis.
	
	\begin{figure}[t!]
		\centering
		\includegraphics[width=.8\linewidth]{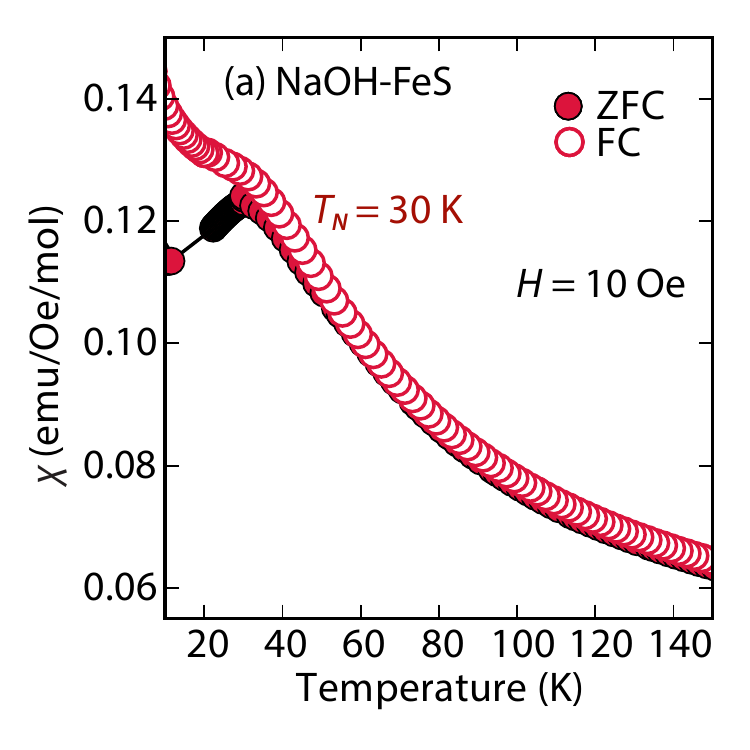}
		\includegraphics[width=.8\linewidth]{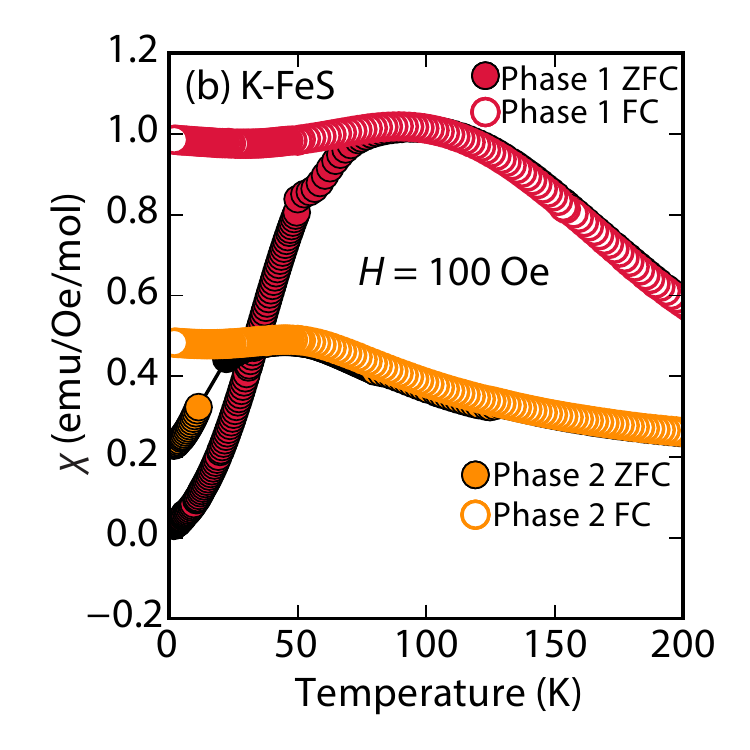}
		\caption{Magnetic susceptibility measurements of (a) Na-tochilinite and (b) K$_x$Fe$_{2-y}$S$_2$, respectively. The lattice constant $c$ for Phase 1 and Phase 2 in (b) is 13.627 and 13.470 \AA, respectively.}
		\label{fig_NaKMPMS}
	\end{figure}
	
	Next, we discuss the nature of the chemical composition of Na-tochilinite. As in some natural minerals, \cite{Pekov2013}  we can formulate the stoichiometry as [(Na$_{1-x}$Fe$_x$)-(OH)$_2$]FeS, and ICP-AES analysis provided an Fe/Na ratio of 2.99. Therefore, [(Na$_{0.5}$Fe$_{0.5}$)(OH)$_2$]FeS is the proposed chemical formula since the ratio of Fe to Na in the tochilinite is (1 + $x$)/(1 - $x$) = 2.99. We can modify the formula by considering the number of iron vacancies in the FeS layers, $y$, and the phase fraction of mackinawite FeS impurity, $f$.   The formula is therefore re-written as (1 + $x - y$)/(1 - $x$) = 2.99$\times$(1 - $f$).  If we estimate the limits based on diffraction data as $y < 0.2$ and $0.1 < f < 0.2$, then $x$ can vary between $0.41<x<0.52$. This result suggests that approximately half of the cations in the hydroxide layers are filled by iron cations, and in order to charge balance the two OH$^{-}$ groups of the hexagonal brucite layer, the nature of that iron site must be in the form of Fe$^{3+}$.  Whilst ICP-AES could not determine the number of hydroxide groups, crystal chemistry arguments support $M$(OH)$_2$ for the spacer layer since this is how the hexagonal brucite is formulated.  Furthermore, the highly reactive and pyrophoric mineral known as ``white rust'' consists of Fe(OH)$_2$ layers that crystallize in the CdI$_2$-type structure.\cite{Parise2000}  By oxidizing to Fe$^{3+}$, such a layer would be stabilized by the presence of either Na$^+$ or vacancies, and indeed natural tochilinites exhibit significant amounts of Fe vacancies (up to 20\%).\cite{Peng2009,Kakos1994} 
	
	Rather than displaying superconductivity as in the LiOH-intercalated systems or long-range ferro- or ferrimagnetism as in $inc$-Na-tochilinite, Na-tochilinite displays broad features in the magnetization reminiscent of short-range antiferromagnetic behavior (Fig. \ref{fig_NaKMPMS}a). The splitting of the ZFC and FC curves likely indicate some degree of spin glassiness. The presence of iron vacancies and distortion of the iron square sublattice are some the likely reasons that Na-tochilinite does not produce a well-defined transition in the magnetization data. Interestingly, Parise \textit{et al.} found through neutron diffraction that Fe(OH)$_2$ exhibits long-range magnetic ordering with each sheet consisting of ferromagnetically coupled iron centers, and each sheet anti-aligned to each other.\cite{Parise2000}  Future neutron diffraction experiments on both incommensurate and quasi-commensurate Na-tochilinites would reveal the nature of the interesting evolution of long-range magnetic ordering arising from the hydroxide layer.

	\subsection*{K-intercalated phases}
	
	Efforts to incorporate KOH layers into FeS hosts resulted in cationic K$^{+}$ intercalation instead. When hydrothermal reactions of Fe powder with KOH and a sulfide source were undertaken, the XRD pattern revealed a phase pure sample similar to the K$_x$Fe$_{2-y}$S$_2$ prepared using solid-state routes (Fig. \ref{fig_XRD_NaKOH}b). In addition, its pattern could be well fit by Rietveld refinement using the crystal structure of K$_x$Fe$_{2-y}$S$_2$ with the space group $I4/mmm$. Its layer spacing (lattice constant $c$) is 13.47 \AA, which is comparable to the reported 245-type ($I4/m$, 13.599 \AA)\cite{Lei2011} and 122-type ($I4/mmm$, 13.546 \AA) compounds.\cite{Ying2012} EDS analysis gave a composition of K$_{1.1}$Fe$_{1.6}$S$_2$ and its magnetic susceptibility displayed broad antiferromagnetic features around 45 K ($c$ = 13.470 \AA) and 96 K ($c$ = 13.627 \AA) for samples with different layer spacings (Fig. \ref{fig_NaKMPMS}b).  The ZFC and FC curves do not trace each other as well, which raises the possibility that these materials may display some spin glassiness as well. Since the transitions are fairly broad, it is likely that long-range antiferromagnetic ordering is never observed but rather some form of low-dimensional or short-range antiferromagnetic order.  Although not superconducting, it is remarkable that we could prepare via hydrothermal routes such ternary phases since these have previously been prepared only by high temperature solid state techniques.
	
	\section*{Conclusions}
	
	In conclusion, we have demonstrated that metal hydroxides can be intercalated into tetragonal mackinawite-type FeS via hydrothermal routes, and that new superconductors can be prepared in this manner.  Given that FeS is a metastable phase, it is of paramount importance that we continue to explore novel low temperature routes towards mineral-inspired superconductors.  Whilst we have enhanced $T_c$ to 8 K through these charge-doping hydroxide layers, we have also demonstrated that FeS can serve as a suitable host for various guests species acting as bases. The differences in going from Li$^{+}$ to Na$^{+}$ to K$^{+}$ are remarkable in the vastly different structure types that were stabilized and the physical properties that are manifested.  These results point to the exciting possibility of utilizing both size and charge parameters of other guests species, such as amines, to ultimately enhance the superconductivity of sulfide-based materials.  Furthermore, the fact that heterostructures could be stabilized points to mackinawite-type FeS as a possible new 2D chalcogenide to be incorporated into other functional 2D materials.  The field of vertical 2D heterostructures has exciting possibilities for constructing entirely new functional materials,\cite{Lotsch2015} and mackinawite-type FeS could be a new building block in such structures.
	
	\section*{Acknowledgements}
	
	We thank Dr. Igor Pekov at Lomonosov Moscow State University for discussions with mineral ferrotochilinite. Research at the University of Maryland was supported by the NSF Career DMR-1455118 and the AFOSR Grant No. FA9550-14-1-0332. Use of the Advanced Photon Source at Argonne National Laboratory was supported by the U. S. Department of Energy, Office of Science, Office of Basic Energy Sciences, under Contract No. DE-AC02-06CH11357.  We also acknowledge support from the the Maryland Nanocenter and Center for Nanophysics and Advanced Materials.

\newpage

    \begin{center}
    	\par \textbf{{\LARGE Supplementary Information}}
    \end{center}
    \vspace{1cm}
    \renewcommand{\thefigure}{S\arabic{figure}}
    \renewcommand{\thetable}{S\arabic{table}}
    
    \setcounter{figure}{0}
    \setcounter{table}{0}
 
 \section*{Diffraction, transport, magnetization and other characterization measurements}
 
 Powder X-ray diffraction (XRD) data were collected using a Bruker D8 X-ray diffractometer with Cu K${\alpha}$ radiation, ${\lambda}$ = 1.5418 \AA. High-resolution synchrotron X-ray diffraction were carried out at Beamline 11-BM at the Advanced Photon Source (APS). Diffraction data were collected between 0.5$^\circ$ and 46$^\circ$ with a step size of 0.0001$^\circ$ using a constant wavelength ${\lambda}$ = 0.414164 \AA \ (30 keV). Rietveld and Pawley refinements were carried out using TOPAS software. Microscopic images were examined on a Hitachi SU-70 SEM field emission scanning electron microscope (SEM), and their elemental compositions were determined by energy dispersive X-ray spectroscopy (EDS) using a BRUKER EDS detector. Electron diffraction patterns were obtained using a JEM 2100 LaB$_6$ transmission electron microscope (TEM) at an acceleration voltage of 200 KeV. 
 
 Inductively coupled plasma atomic emission spectroscopy (ICP-AES) data were collected using an Shimadzu ICPE-9000 spectrometer. Standards used for ICP-AES were diluted from 1000 ppm of respective elements purchased from Sigma-Aldrich. 
 
 Magnetic susceptibility measurements were performed using a Quantum Design Magnetic Properties Measurement System (MPMS). The volume fractions of superconducting phases were calculated based on the density obtained from Reitveld refinement. Electrical resistivity and heat capacity measurements were performed on a 14 T Quantum Design Physical Properties Measurement System (PPMS).

 \section*{More notes on the K$_x$Fe$_{2-y}$S$_2$ phase}

 While we did not find superconducting phases containing potassium, we did demonstrate that the synthetic temperature for the preparation of K${_x}$Fe${_{2-y}}$S${_2}$ can be lowered from about 1000 $^{\circ}$C to 160 $^{\circ}$C through hydrothermal methods.  Without KOH, single crystals of K${_x}$Fe${_{2-y}}$S${_2}$ can be completely converted to mackinawite FeS.  Therefore, the conversion between K${_x}$Fe${_{2-y}}$S${_2}$ and tetragonal FeS is fully reversible, as traced by the equilibrium reaction between $\textbf{1}$ and $\textbf{2}$ in Fig. 1. With further work on reducing the iron vacancies, the potassium intercalated phases could be made superconducting. To confirm this, we started to apply this route to the selenide system without optimization, and K${_x}$Fe${_{2-y}}$Se${_2}$ was prepared despite the presence of tetragonal FeSe. The implication of these results are that this hydrothermal route can lead to pure 122 type of layered compounds or the corresponding deintercalated tetragonal system. In addition, this hydrothermal route can be advantageous over solid-state route to avoid high temperature impurity phases or targeting compounds not thermodynamically stable at low temperature. 
 
 \section*{More notes on the structure of Na-Tochinilite}
 
 A projection of Fe atoms on the (001) plane in Na-tochilinite is illustrated in Fig. \ref{fig_Toch}, and compared to perfect square lattice in tetragonal FeS, there is a clear distortion along the $b$-axis.
 
 The comparison between the morphologies of (Li$_{1-x}$Fe$_x$OH)FeS and NaOH intercalated FeS systems may provide further evidence of the hexagonal hydroxide layers with the NaOH reactions. Our (Li$_{1-x}$Fe$_x$OH)FeS samples consisted mainly of square-shaped platelets in micron size (Fig. \ref{fig_SEM}a and Fig. \ref{fig_SEM}b), which would be indicative of the underpinning layered tetragonal structure. However, a similar morphology was not observed for either the $inc$-Na-tochinilite or  Na-tochinilite samples(Fig.\ref{fig_SEM}c and Fig. \ref{fig_SEM}d, respectively). While still layered, the crystallites display irregular shapes instead of square platelets. The square-shaped platelets are consistent with the crystal habit of the tetragonal LiOH-intercalated FeS system. 
 
 \scriptsize{
 	\bibliography{AOHFeS_manuscript} 

 \begin{table}[]
 	\centering
 	\caption{List of Li${_{1-x}}$Fe$_{x}$OH)FeS samples. Detailed synthetic conditions are described in the above text, and only temperature, the most important factor, is shown in the table. Lattice constants of only representative samples are shown for duplicate samples. Because Na-tochilinite can be produced with the presence of NaOH, Na$_{2}$S ${\cdot}$ ${9}$H$_{2}$O was not used as a precursor for powder samples due to its hydrolysis to NaOH and NaSH in solution. Li$_2$S was the main sulfur source used for powder samples, and every sample prepared with Li$_2$S has been reproduced at least once. Single crystal samples are not very homogeneous, and their $T_c$'s vary from crystal to crystal, but their superconductivity is highly reproducible. Multiple single crystal batches have been reproduced at 120 $^\circ$C suing different sulfur sources with the presence of Sn.}
 	\label{Table_samples}
 	\resizebox{\columnwidth}{!}{
 		\begin{tabular}{llllllll}
 			\hline
 			\hline
 			& \textbf{No.}  & \textbf{Temperature ($^\circ$C) }  & \textbf{Sulfur source} & Sn (Y/N) & \textbf{$T_c$ (K)} &  \textbf{$a$ (\AA)}& \textbf{$c$ (\AA)}  \\
 			\hline
 			\multirow{10}{*}{Powder}        & 1  & 130  & Li$_2$S & N  & N/A  & 3.706  & 8.862 \\
 			& 2  & 160  & Li$_2$S & N  & N/A  & 3.704  & 8.942 \\
 			& 3  & 160  & thiourea & N  & N/A  & 3.696  & 8.979  \\
 			& 4  & 180  & thiourea & N  & N/A  & 3.702  & 8.943  \\
 			& 5  & 200  & thiourea & N  & N/A  & 3.702  & 8.970 \\
 			& 6  & 120  & thiourea & Y  & 2-3  & 3.700  & 8.919  \\
 			& 7  & 120  & Li$_2$S  & Y  & 2-3  & 3.704  & 8.900  \\
 			& 8  & 140  & Li$_2$S  & Y  & 2-3  & 3.706  & 8.900  \\
 			& 9  & 160  & Li$_2$S  & Y  & 2-3  & 3.704  & 8.888  \\
 			& 10 & 200  & Li$_2$S  & Y  & N/A  & 3.701  & 8.926  \\
 			\hline
 			\multirow{3}{*}{SC} & 11  & 120  & thiourea & Y  & 2-8  & 3.703  & 8.935  \\
 			& 12  & 120  & Na$_{2}$S ${\cdot}$ ${9}$H$_{2}$O & Y  & 2-6  & 3.712  & 8.877  \\
 			& 13  & 120  & Li$_2$S & Y  & 2-4  & 3.703  & 8.960 \\
 			\hline
 			\hline
 		\end{tabular}}
 	\end{table}
 	
 	\begin{figure}[h]
 		\centering
 		\includegraphics[width=.45\linewidth]{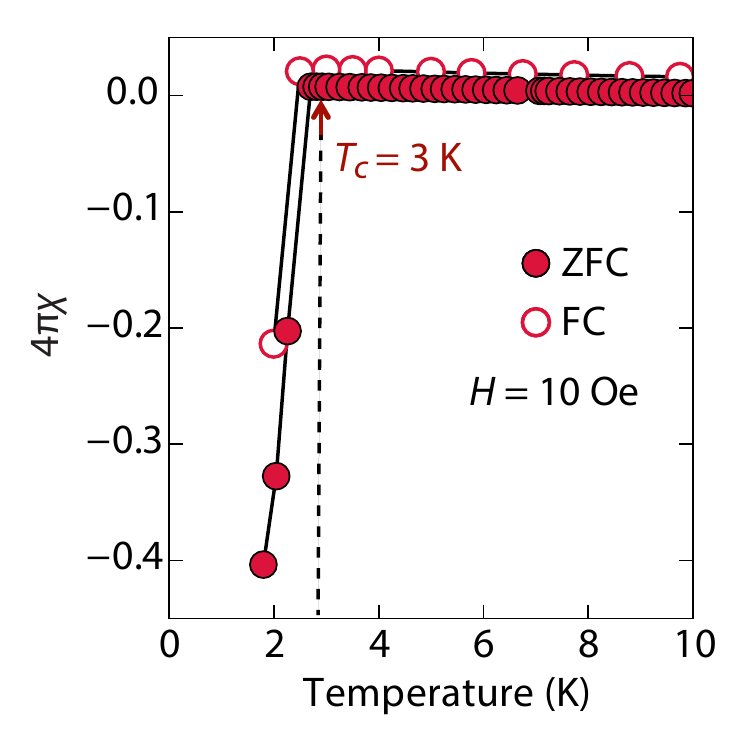}
 		\includegraphics[width=.45\linewidth]{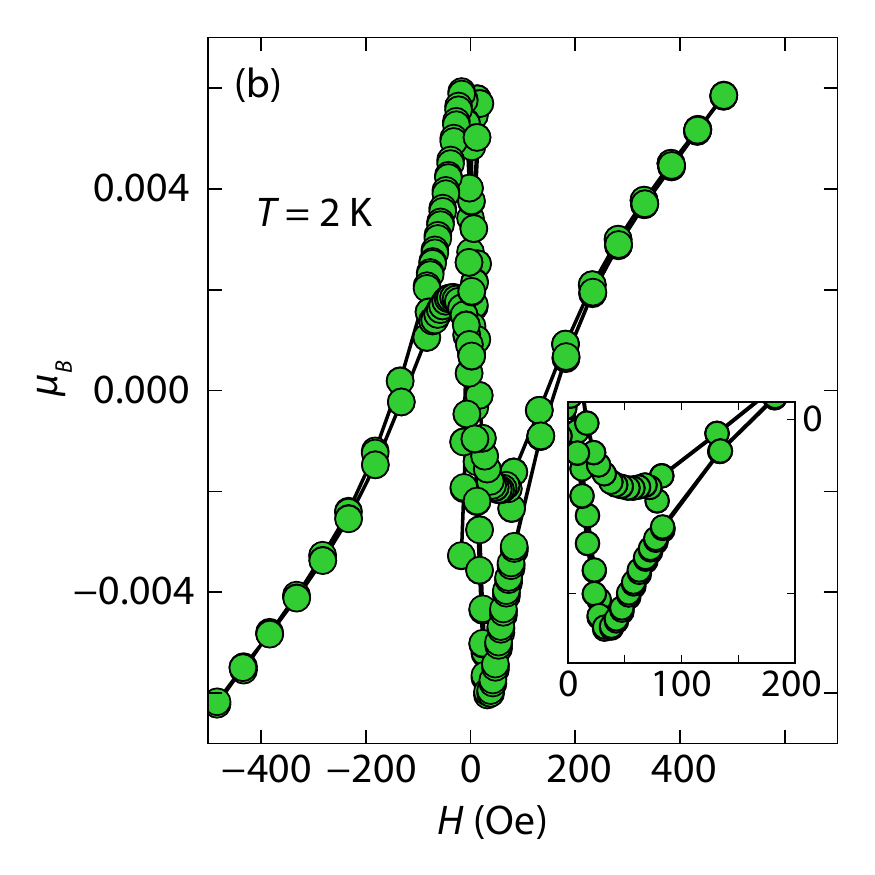}
 		\caption{Magnetic susceptibility measurements of superconducting (Li$_{1-x}$Fe$_x$OH)FeS at (a) constant field and (b) constant temperature. The $H_{c1}$ and $H_{c2}$ of this sample are about 40 and 180 Oe, respectively. The XRD pattern of this sample is shown in Fig. 3a}
 		\label{fig_079b}
 	\end{figure}
 	
 	\begin{figure}[h]
 		\centering
 		\includegraphics[width=.45\linewidth]{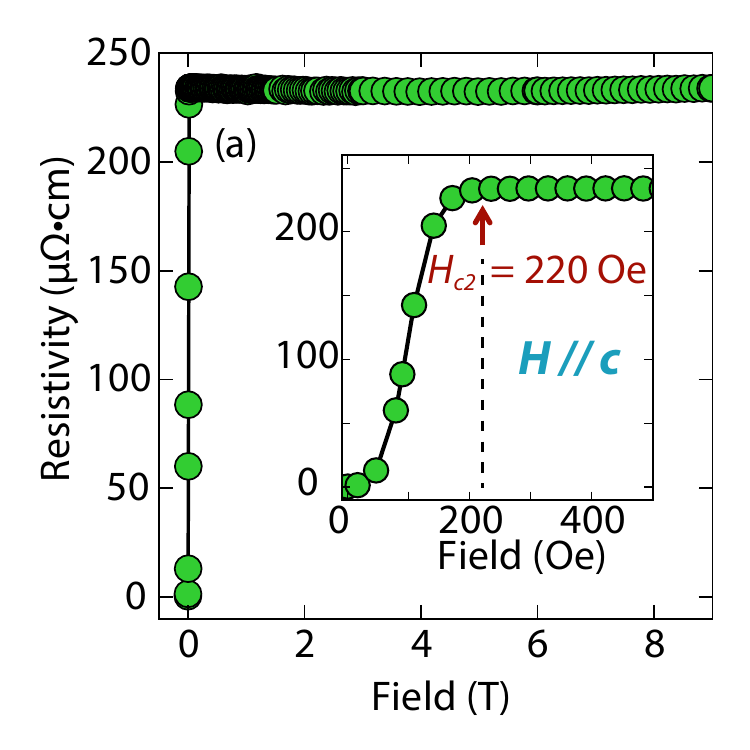}
 		\includegraphics[width=.45\linewidth]{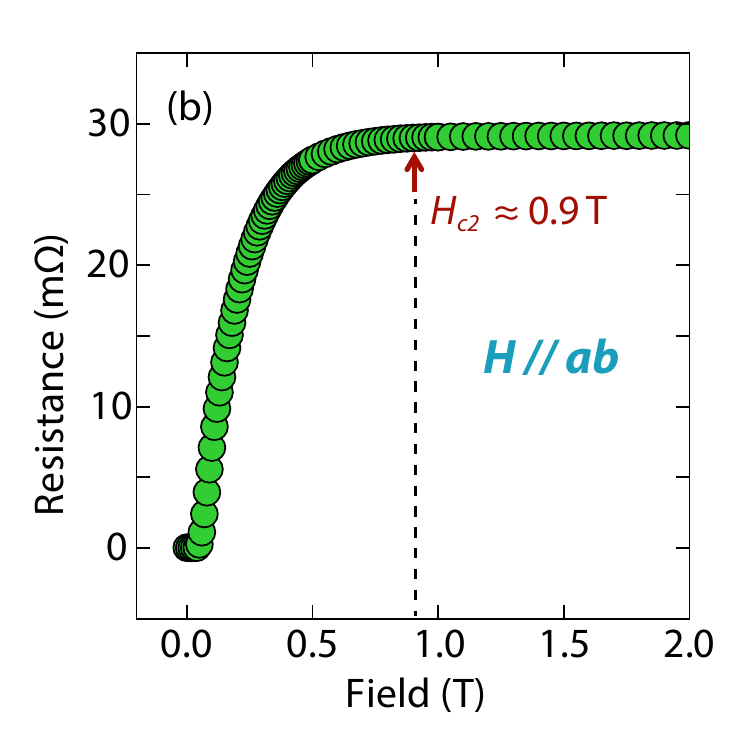}
 		\caption{Field dependence of electrical resistance for a superconducting (Li$_{1-x}$Fe$_x$OH)FeS sample ($T_c$ = 3.5 K) at 1.8 K. The anisotropy of H//c and H//ab are shown in (a) and (b), respectively. Its temperature dependent electrical resistivity is shown in Fig. 2a.}
 		\label{fig_Hc2_LiOH}
 	\end{figure}
 	
 	\begin{figure}[h]
 		\centering
 		\includegraphics[width=.8\linewidth]{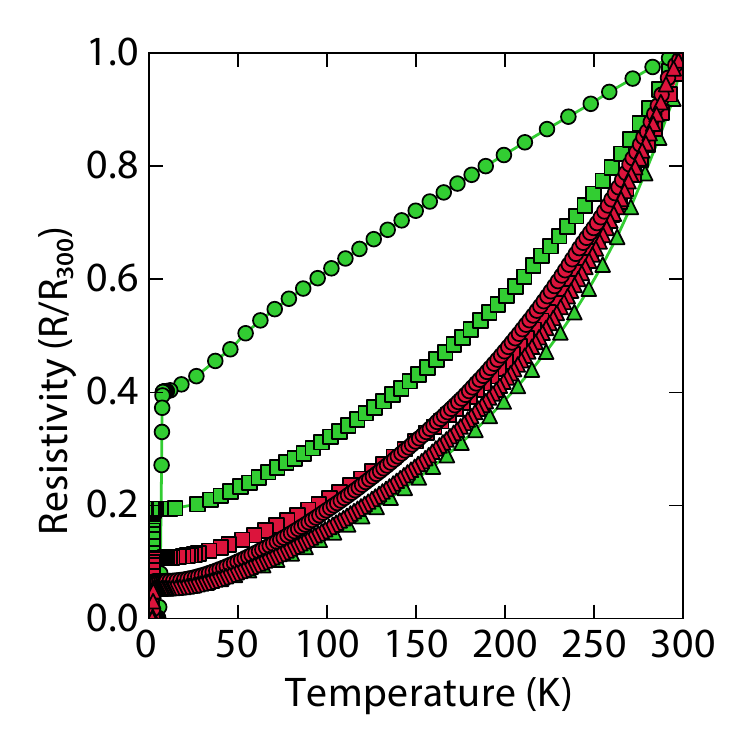}
 		\caption{Temperature dependence of electrical resistivity for superconducting (Li$_{1-x}$Fe$_x$OH)FeS samples. Green and red colors indicate samples prepared using thiourea and Na$_{2}$S ${\cdot}$ ${9}$H$_{2}$O, respectively. }
 		\label{fig_RRR}
 	\end{figure}	
 	\begin{figure}[h]
 		\centering
 		\includegraphics[width=.8\linewidth]{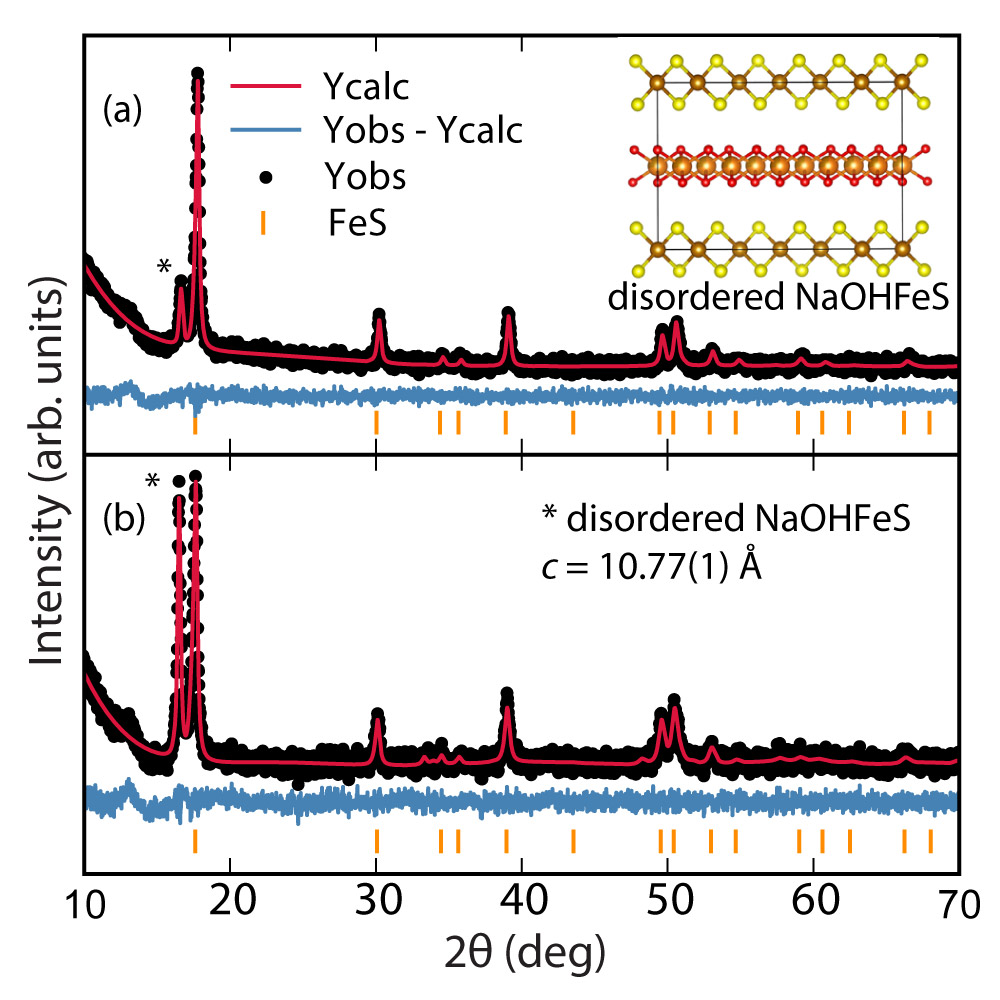}
 		\caption{XRD patterns of mixtures of disordered NaOH intercalated FeS (indicated by *) and tetragonal FeS (indicated by tick marks) with significantly more tetragonal FeS in (a) than (b). The magnetic susceptibility of (a) and (b) are shown in Fig. \ref{fig_XZB050b} and Fig. 4, respectively. }
 		\label{fig_XRD_SCNaOH}
 	\end{figure}	
 	
 	\begin{figure}[h]
 		\centering
 		\includegraphics[width=.8\linewidth]{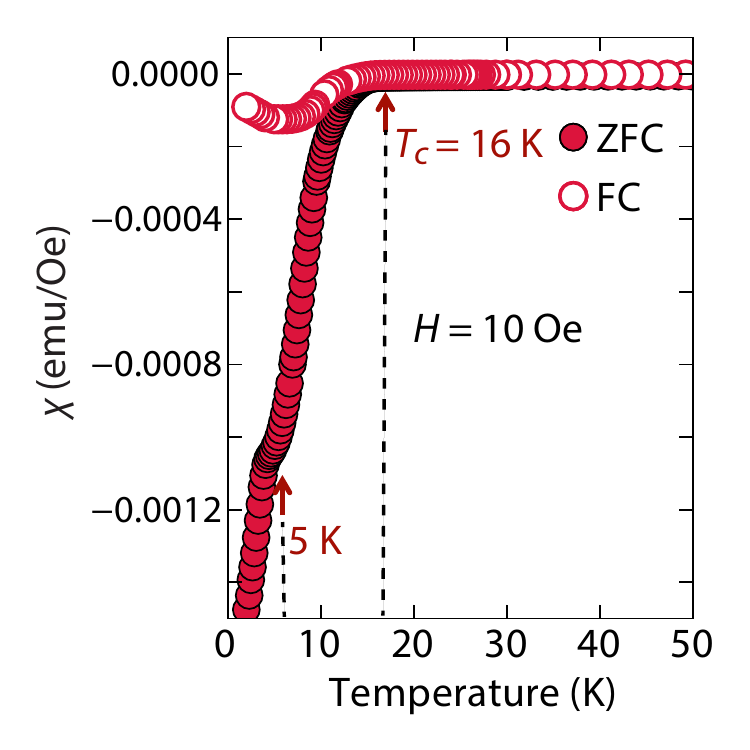}
 		\caption{Temperature dependent magnetic susceptibility measurement of $inc$-Na-tochinilite with tetragonal FeS as a major phase. Its XRD pattern is shown in Fig. \ref{fig_XRD_SCNaOH}a}
 		\label{fig_XZB050b}
 	\end{figure}
 	
 	\begin{figure}[!h]
 		\centering
 		\includegraphics[width=.8\linewidth]{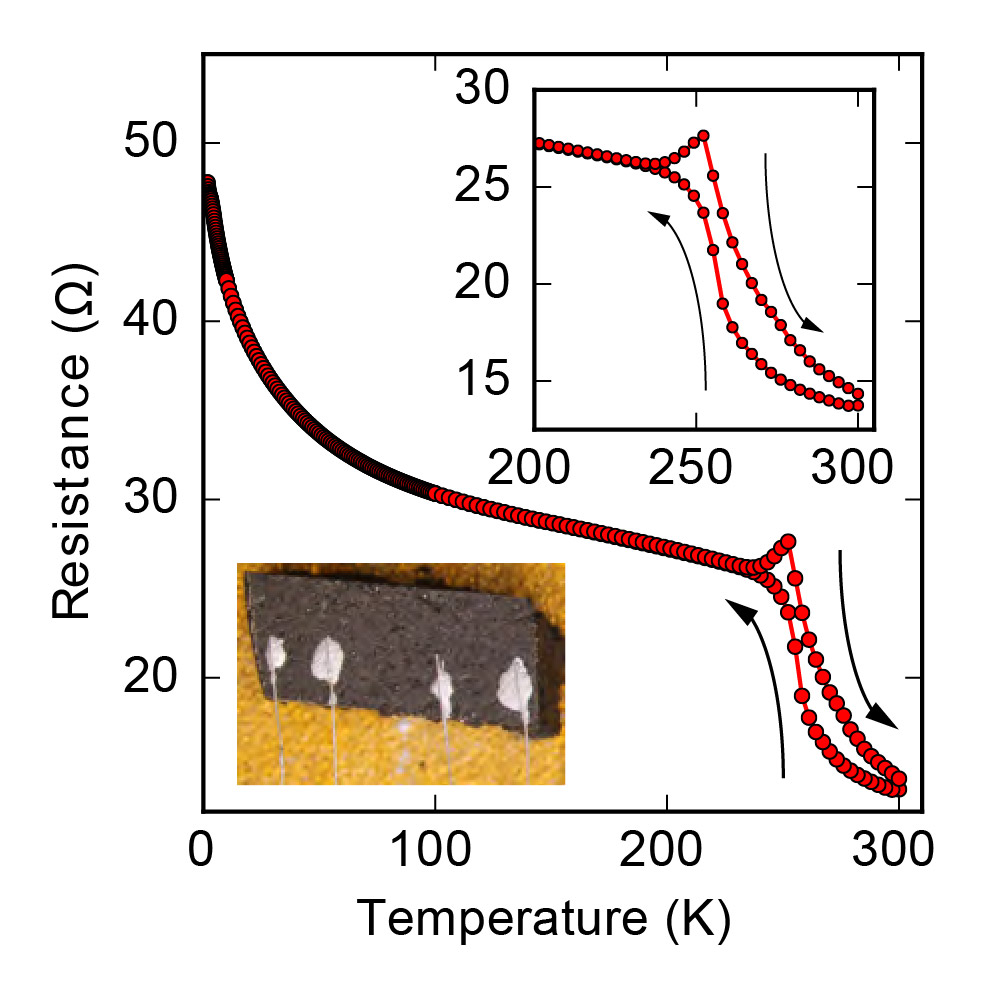}
 		\caption{Electrical resistance of $inc$-Na-tochilinite as a function of temperature. The measurement was carried out on a pressed pellet from powders.}
 		\label{fig_rvT_NaOH}
 	\end{figure}
 	
 	\begin{figure}[!h]
 		\centering
 		\includegraphics[width=.7\linewidth]{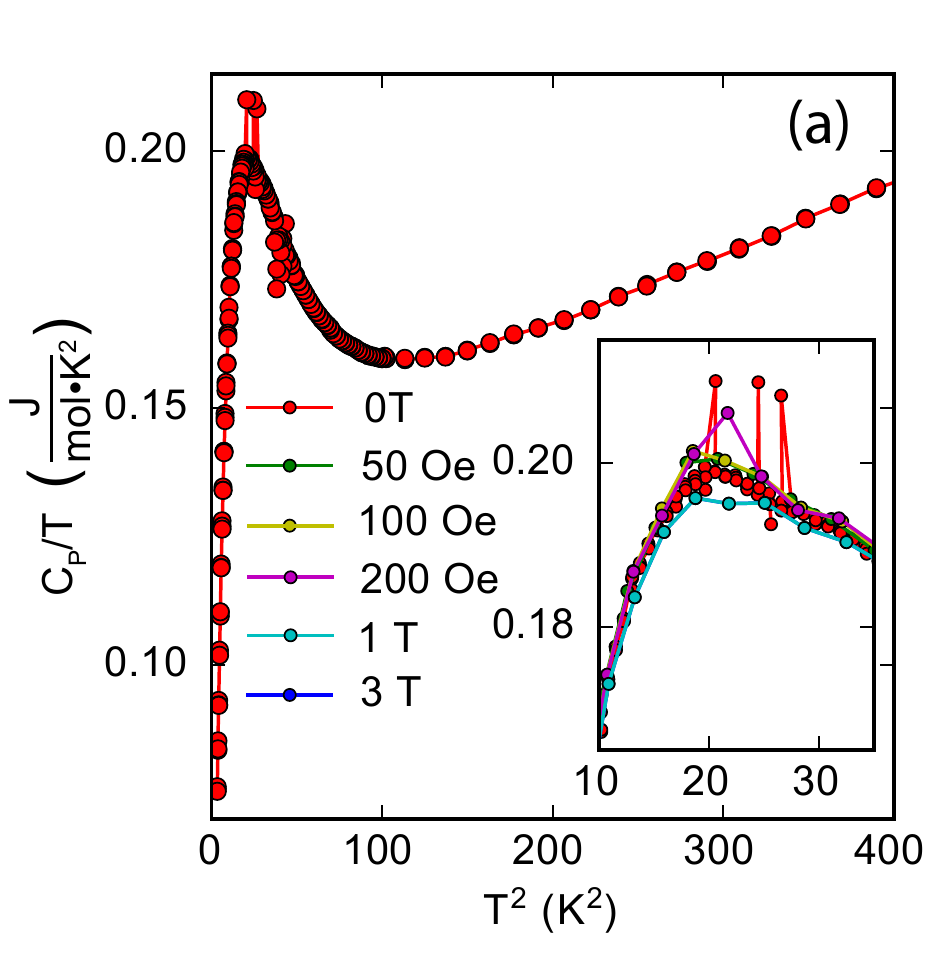}
 		\includegraphics[width=.7\linewidth]{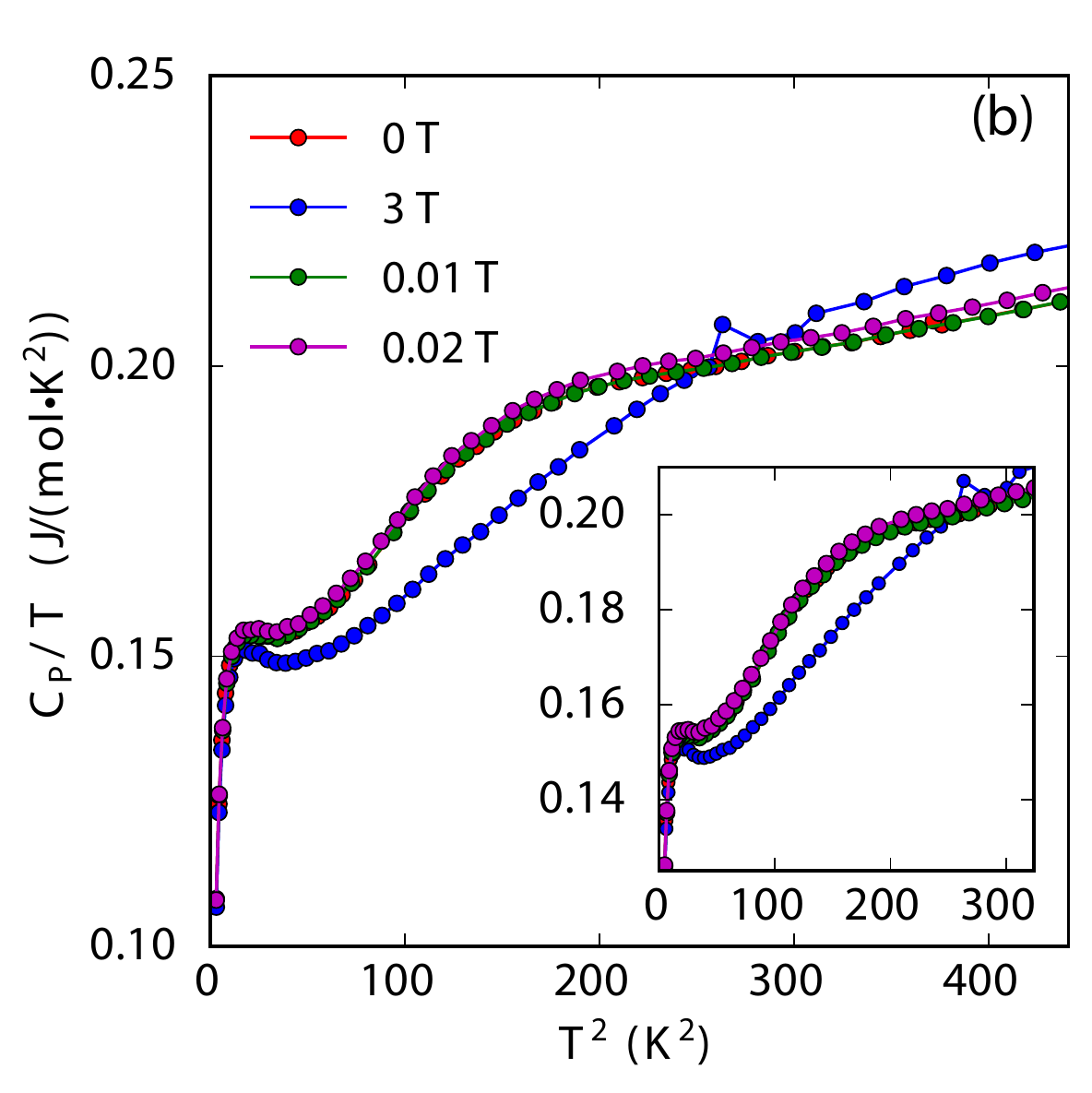}
 		\caption{Specific heat measurements of (a) LiOH-intercalate FeS and (b) $inc$-Na-tochilinite.}
 		\label{fig_LiNaOH_SH}
 	\end{figure}

 	\begin{figure}[h]
 		\centering
 		\includegraphics[width=0.85\columnwidth]{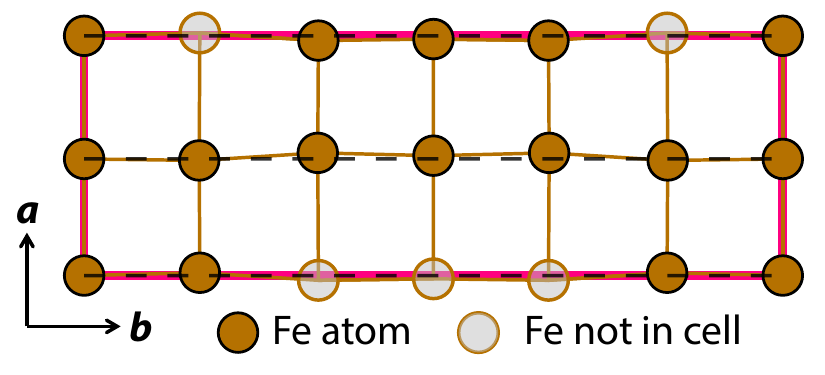}
 		\caption{A projection of the Fe atoms on the (001) plane of the naturally occurring mineral tochilinite, (2(Fe${_{1-x}}$S)${\cdot}$1.8[(Mg, Fe)(OH)$_2$]). A similar distortion of FeS square lattice is observed for Na-tochilinite as suggested by electron diffraction.}
 		\label{fig_Toch}
 	\end{figure}
 	
 	\begin{figure}[h]
 		\centering
 		\includegraphics[width=.8\linewidth]{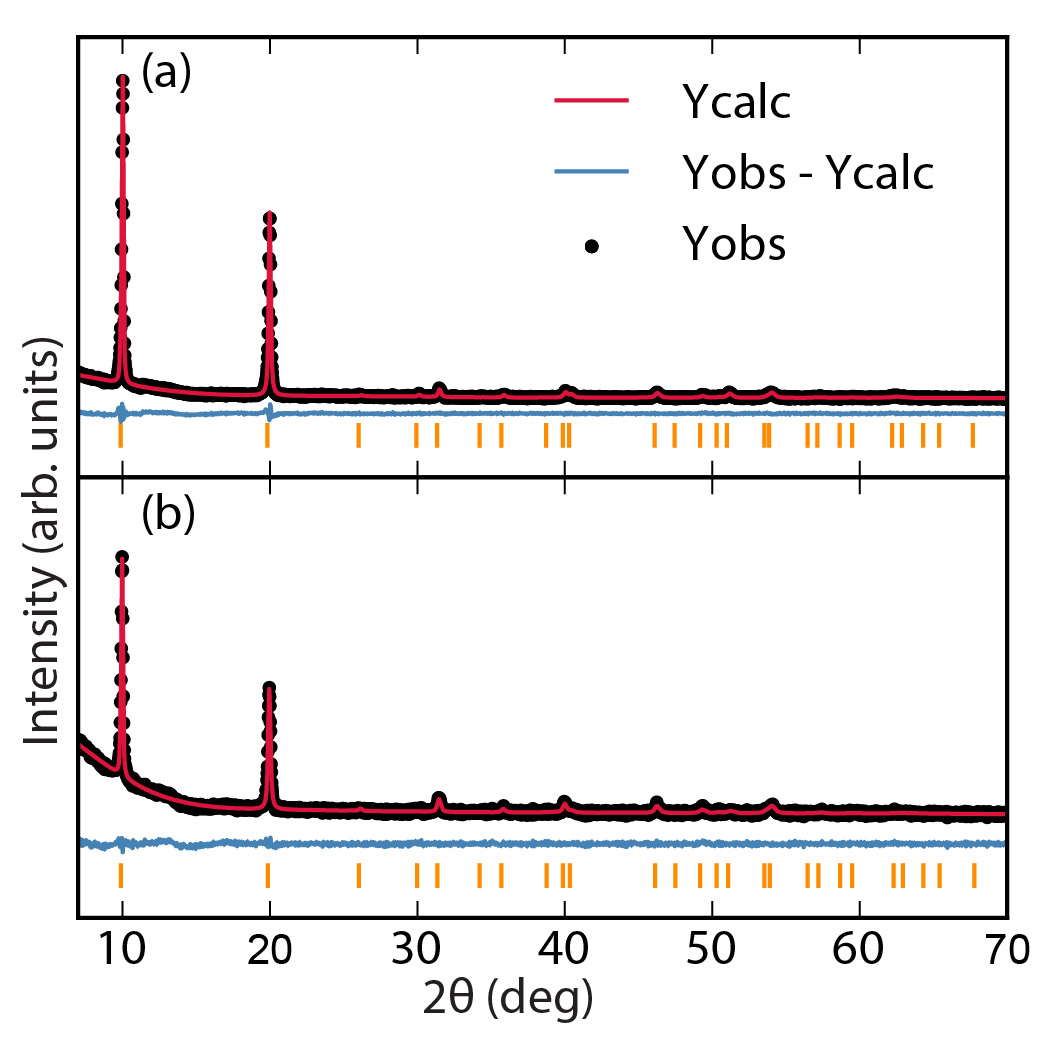}
 		\caption{XRD patterns of LiOH-intercalated FeS samples shown in (a) Fig. 2a and (b) Fig. 2b, respectively. Both are fitted to a $P4/nmm$ space group and show no impurity phases. }
 		\label{fig_XRD_LiOHFeS}
 	\end{figure}

 	\begin{figure}[h]
 		\centering
 		\includegraphics[width=.45\linewidth]{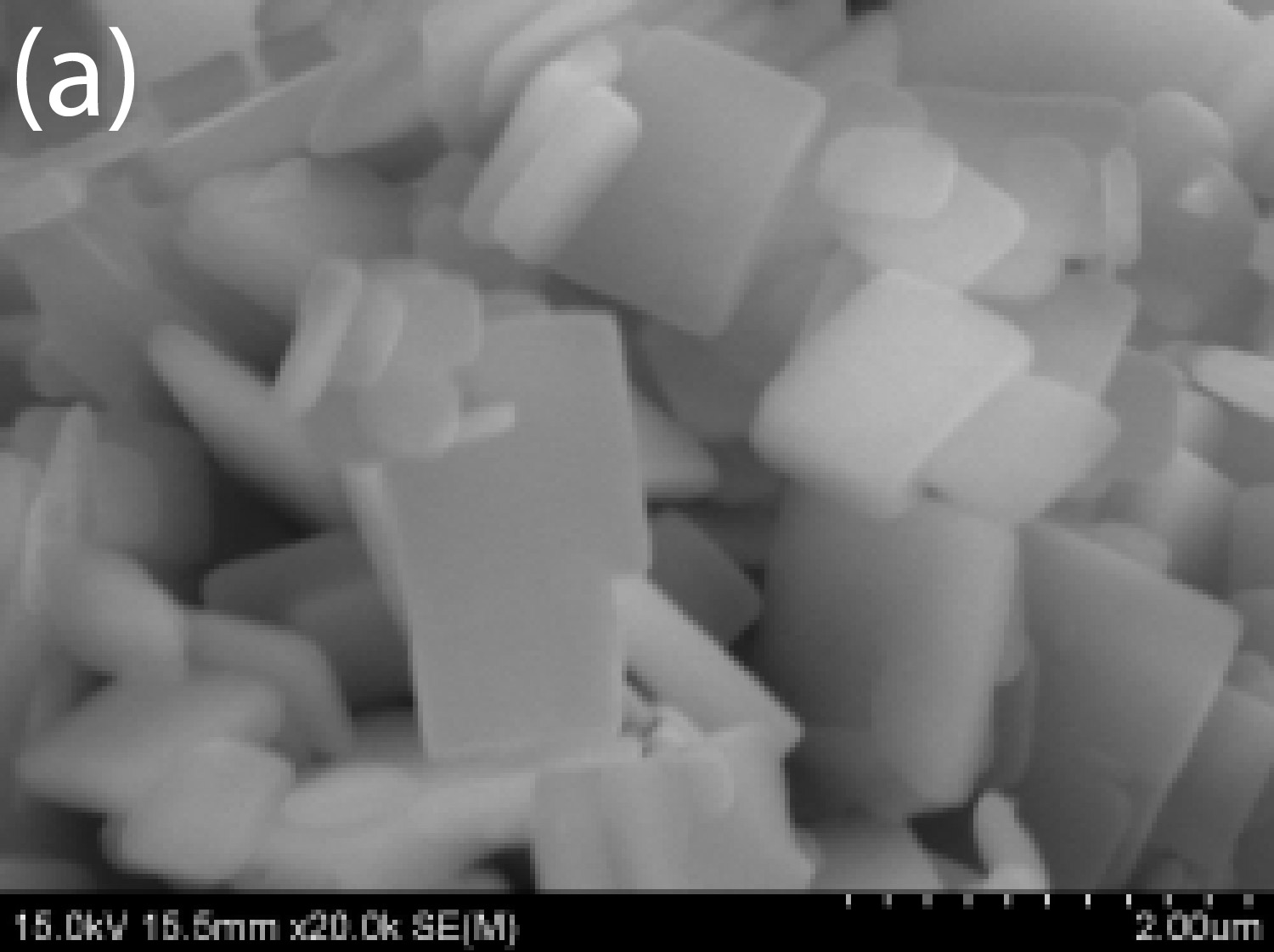}
 		\includegraphics[width=.45\linewidth]{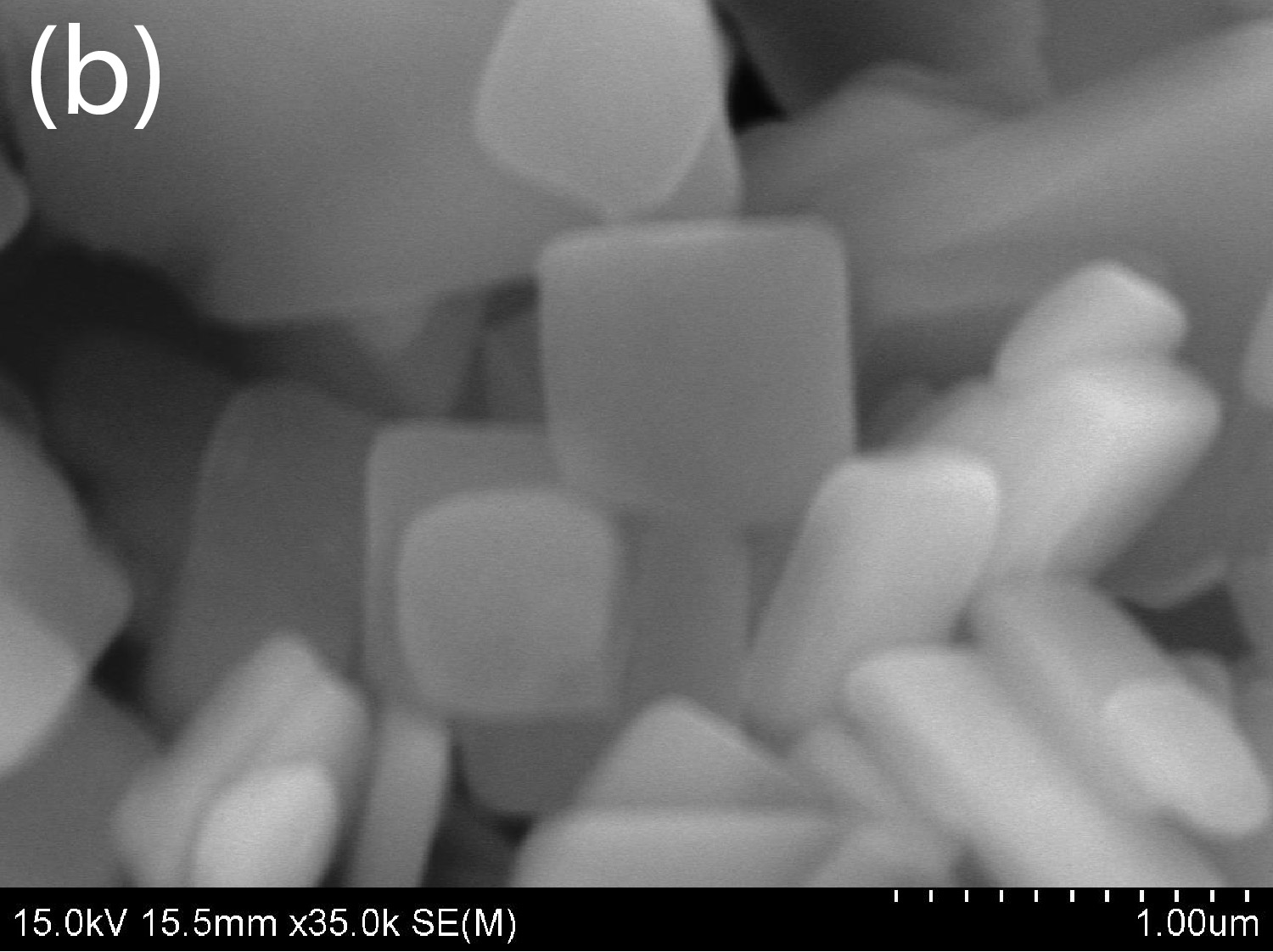}
 		\includegraphics[width=.45\linewidth]{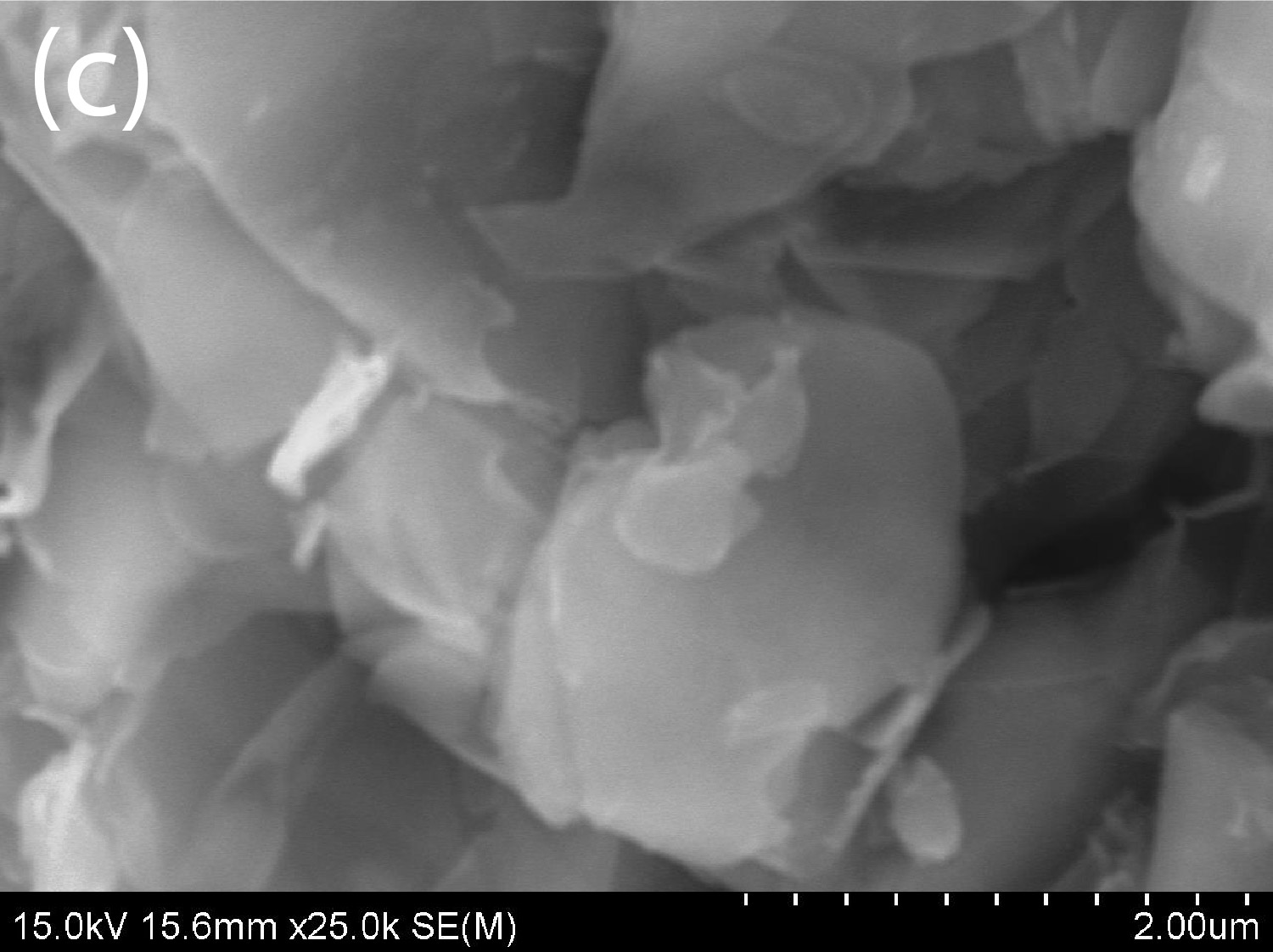}
 		\includegraphics[width=.45\linewidth]{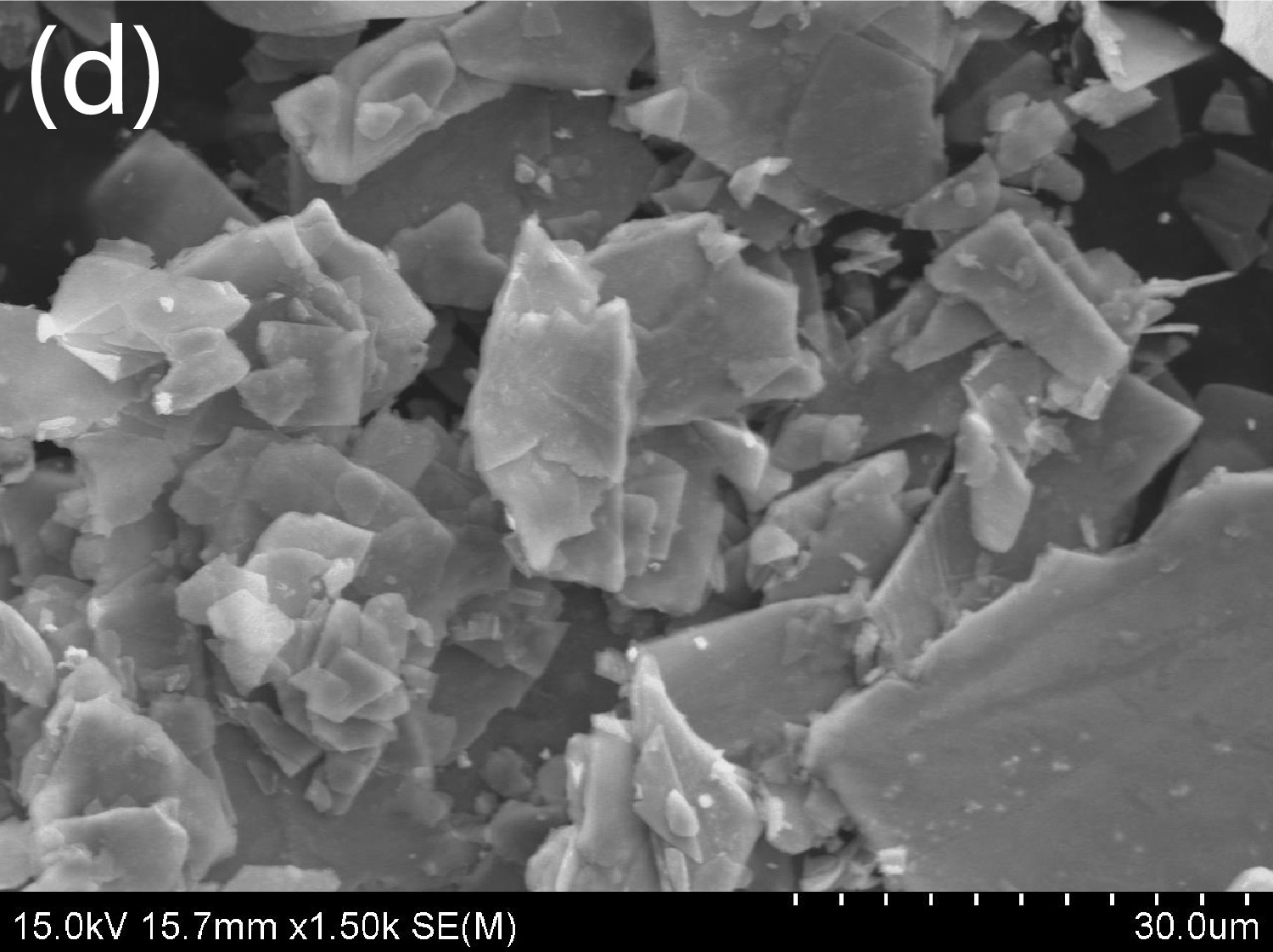}
 		\caption{Scanning electron microscopy (SEM) images of (a) and (b) (Li$_{1-x}$Fe$_x$OH)FeS, (c) $inc$-Na-tochilinite and (d) Na-tochilinite}.
 		\label{fig_SEM}
 	\end{figure}

\end{document}